\documentclass[superscriptaddress,twocolumn,pre]{revtex4-2}
\usepackage{amsmath}
\usepackage{amsfonts}
\usepackage{amssymb}
\usepackage{graphicx}
\graphicspath{{figures/}}
\usepackage{color}
\usepackage[english]{babel}

\usepackage[pdfstartview=FitH,
            breaklinks=true,
            bookmarksopen=false,
            bookmarksnumbered=true,
            colorlinks=true,
            linkcolor=black,
            citecolor=black,
            urlcolor=black,
            pdftitle={},
            pdfauthor={},
            pdfsubject={}
            ]{hyperref}

\newcommand{\ud}{\mathrm{d}}

\begin{document}
\title{Heterologous autoimmunity and prokaryotic immune defense}
\author{Hanrong Chen}
\thanks{These two authors contributed equally to this work.}
\affiliation
{Computational \& Systems Biology, Genome Institute of Singapore, Singapore}
\affiliation{David Rittenhouse Laboratory, Department of Physics and Astronomy, University of Pennsylvania, USA}
\author{Andreas Mayer}
\thanks{These two authors contributed equally to this work.}
\affiliation{Lewis-Sigler Institute for Integrative Genomics, Princeton University, USA}
\author{Vijay Balasubramanian}
\affiliation{David Rittenhouse Laboratory, Department of Physics and Astronomy, University of Pennsylvania, USA}
\affiliation{Theoretische Natuurkunde, Vrije Universiteit Brussel, Belgium}
\date{\today}

\begin{abstract}
Some prokaryotes possess CRISPR-Cas systems that provide adaptive immunity to viruses guided by DNA segments called spacers acquired from invading phage. However, the patchy incidence and limited memory breadth of CRISPR-Cas systems suggest that their fitness benefits are offset by costs. Here, we propose that cross-reactive CRISPR targeting can lead to heterologous autoimmunity, whereby foreign spacers guide self-targeting in a spacer-length dependent fashion. Balancing antiviral defense against autoimmunity predicts a scaling relation between spacer length and CRISPR repertoire size. We find evidence for this scaling through comparative analysis of sequenced prokaryotic genomes, and show that this association also holds at the level of CRISPR types. In contrast, the scaling is absent in strains with nonfunctional CRISPR loci. Finally, we demonstrate that stochastic spacer loss can explain variations around the scaling relation, even between strains of the same species. Our results suggest that heterologous autoimmunity is a selective factor shaping the evolution of CRISPR-Cas systems.
\end{abstract}

\maketitle

Clustered regularly interspaced short palindromic repeat (CRISPR) loci and CRISPR-associated (Cas) proteins form a prokaryotic immune defense \cite{Barrangou2007}. CRISPR loci comprise DNA repeats alternating with variable DNA segments called spacers, which are acquired from invading phage and other foreign genetic material.
Spacer RNA, generated by
CRISPR locus expression and processing, 
guides Cas proteins in a process called interference to bind and cleave target DNA in a sequence-specific manner. Thus, spacers acquired during phage attack confer adaptive, heritable resistance to subsequent invasions.

Different CRISPR-Cas systems use divergent Cas proteins and distinct mechanisms to perform each defense stage \cite{Makarova2019}. For example, spacer acquisition is mediated by the Cas1--Cas2 adaptation module, which sets spacer lengths within a range varying by system \cite{Wang2015,Nunez2015}.
Likewise, CRISPR array sizes range from less than 10 to several hundred spacers, and the full repertoire of a prokaryotic host may comprise several arrays \cite{Pourcel2020}.
Broad repertoires confer resistance to a greater diversity of phages and escape mutants \cite{van_houte_diversity-generating_2016}, but increase constitutive costs of Cas protein expression \cite{vale_costs_2015}, prevent horizontal transfer of beneficial mobile genetic elements \cite{marraffini_crispr_2008,jiang_dealing_2013}, and yield diminishing returns due to finite Cas copy numbers \cite{Martynov2017,Bradde2020}.
CRISPR-Cas systems also cause autoimmunity when spacers guide interference of the host genome, causing mutational pressure in the CRISPR-cas locus and target region \cite{Stern2010,Vercoe2013,Paez-Espino2013,Wei2015}. The patchy incidence of CRISPR in prokaryotes (40\% of bacteria and 85\% of archaea \cite{Makarova2019}), and diverse mechanisms for self--nonself discrimination \cite{Marraffini2015}, suggest that avoiding autoimmunity is a constraint in CRISPR-Cas evolution \cite{Vercoe2013,Paez-Espino2013,Wei2015,Edgar2010,Marraffini2015,Goldberg2018,Makarova2019,Rollie2020}.

Several mechanisms suppress autoimmunity arising from different forms of self-targeting \cite{Marraffini2015}. In type I and II CRISPR systems, interference requires protospacer-adjacent motifs (PAMs), 2--5-nt-long sequences adjacent to target DNA but absent in CRISPR repeats, preventing interference within the CRISPR array \cite{Deveau2008,Semenova2011}. In type III systems, interference requires transcription of target DNA, which avoids targeting phages integrated into the host chromosome (prophage) unless they are actively expressed \cite{Goldberg2014}. Spacers acquired from the host genome are naturally self-targeting, but there are mechanisms to suppress such acquisition \cite{Levy2015,Modell2017}. For example, type I-E systems acquire spacers preferentially at double-stranded DNA breaks, which occur primarily at stalled phage DNA replication forks, and acquisition is confined by Chi sites which are enriched in bacterial genomes \cite{Levy2015}.

Here, we propose that foreign spacer acquisition is also a source of autoimmunity, an effect we term {\it heterologous autoimmunity}. This occurs if a foreign spacer and a segment of the host genome are sufficiently similar, the likelihood of which depends on sequence statistics and CRISPR targeting specificity. Heterologous autoimmunity resembles off-target effects in CRISPR-Cas genome editing \cite{Fu2013,Hsu2013}, but the implications for adaptive immunity have not been explored.
We combine probabilistic modeling with comparative analyses of CRISPR repertoires to show that: (a) heterologous autoimmunity is a significant threat, (b) avoidance of autoimmunity predicts a scaling relation between spacer length and CRISPR repertoire size, (c) the scaling is empirically present across prokaryotes, (d) proportionate use of different CRISPR systems underpins the scaling, and (e) stochastic spacer loss can explain variability in individual strains around the scaling relation.
Our work suggests that avoidance of heterologous autoimmunity is a force shaping the evolution of CRISPR-Cas systems.

\begin{figure*}[t]
    \includegraphics[width=\linewidth]{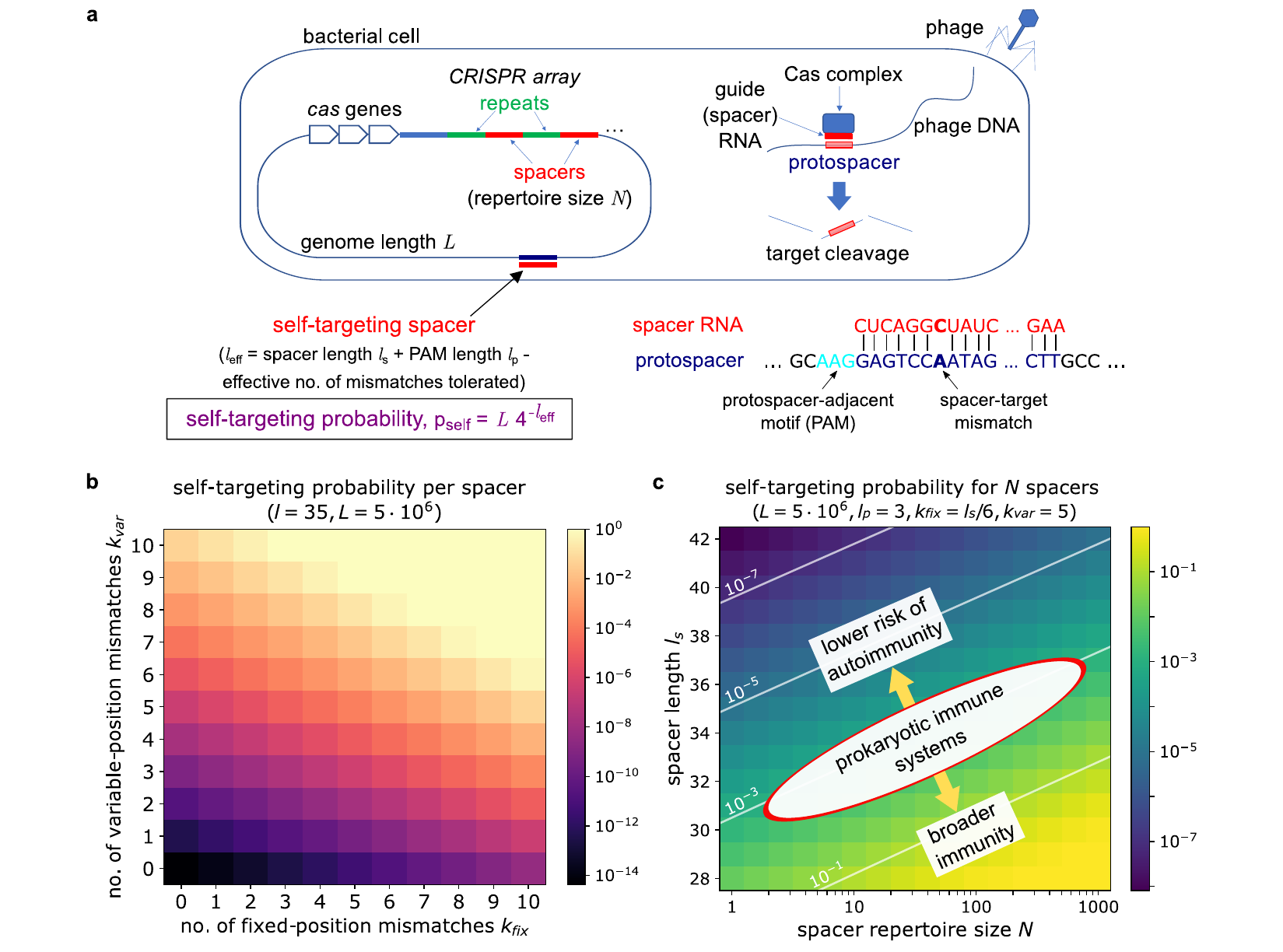}
    \caption{{\bf CRISPR-Cas immune defense incurs a risk of heterologous autoimmunity.}
    {\bf a}, Sketch of the main components of CRISPR-Cas immune defense.
    {\bf b}, Per-spacer probability of heterologous self-targeting, $p_{\text{self}}$, as a function of the number of tolerated mismatches at fixed and variable positions along the spacer, $k_{\text{fix}}$ and $k_{\text{var}}$, respectively.
    {\bf c}, We hypothesize that CRISPR-Cas systems incur a risk of heterologous autoimmunity. The probability that this occurs depends strongly on spacer length, implying a scaling between repertoire size and spacer length.
    }
    \label{fig:model}
\end{figure*}

\section{Results}

\subsection{Cross-reactivity leads to autoimmunity}

We treat heterologous self-targeting as a sequence-matching problem \cite{Percus1993,karlin_methods_1990,dembo_limit_1994},
and derive estimates for the probability of a spacer being sufficiently similar to at least one site in the host genome.
For a spacer of length $l_s$ and PAM of length $l_p$ (where it exists), an exact match at a given position requires $l \equiv l_s + l_p$ complementary nucleotides. In a host genome of length $L$, where $L \gg l$, there are $L - l + 1 \approx L$ starting positions for a match. At leading order, and ignoring nucleotide usage biases, we may treat matches as occurring independently with probability $4^{-l}$. Thus, the probability of an exact match anywhere on the genome is (see Methods)
\begin{equation}
    p_0 \equiv L 4^{-l}, \text{ where } l = l_s + l_p.
    \label{eq:p_0}
\end{equation}
Using order-of-magnitude estimates for the E. coli type I-E system ($L = 5 \times 10^6$ nt, $l_s = 32$ nt, and $l_p = 3$ nt) gives a negligible probability $p_{\text{0}} \sim 10^{-15}$.

However, CRISPR interference tolerates position- and nucleotide-dependent mismatches between spacer RNA and target DNA \cite{Fu2013,Hsu2013,Fineran2014,Jung2017}. In general, mismatches in the PAM are not allowed, and mismatches in  PAM-distal regions are more tolerated than mismatches in the PAM-proximal region \cite{Semenova2011}. Up to $\sim 5 $ mismatches are allowed in type II systems \cite{Fu2013,Hsu2013}, while in type I-E systems, errors are mostly tolerated at specific positions with 6-nt periodicity \cite{Fineran2014,Jung2017}. Furthermore, partial spacer-target matching may also trigger priming, which is the rapid acquisition of new spacers from regions surrounding target DNA \cite{Datsenko2012,Fineran2014,Staals2016}. In type I-E and I-F systems, priming can be triggered despite as many as 10 mismatches in the PAM and target region \cite{Fineran2014,Staals2016}. Thus, a foreign spacer that partially matches a region of the host genome, while not causing direct interference, may still trigger primed acquisition of self-spacers \cite{Staals2016} and hence cause autoimmunity.

Since CRISPR interference specificity and primed acquisition have only been characterized for a few systems,  we consider here two general classes of mismatch tolerance including the above cases: (a) mismatches at $k_{\text{fix}}$ fixed positions, and (b) mismatches at $k_{\text{var}}$ variable positions anywhere in the target region. These increase the per-spacer self-targeting probability by a combinatorial factor  $\alpha(k_{\text{fix}}, k_{\text{var}},l)$ (see Methods):
\begin{equation}
    p_{\text{self}} = \alpha(k_{\text{fix}}, k_{\text{var}},l) \, p_0  = 
    \alpha(k_{\text{fix}},k_{\text{var}},l) \, L \, 4^{-l}.
    \label{eq:p_self}
\end{equation}

A greater number of allowed mismatches greatly increases the likelihood of heterologous self-targeting (Fig. \ref{fig:model}b).
To gain intuition we rewrite Eq.~\ref{eq:p_self} as
\begin{gather}
    p_{\text{self}} \equiv L \,  4^{-l_{\text{eff}}}, \text{ where}     \label{eq:l_eff}  \\
    l_{\text{eff}}(l,k_{\text{fix}},k_{\text{var}}) \equiv  l - \log_4\alpha \approx l - k_{\text{fix}} - k_{\text{var}}\log_4 3(l - k_{\text{fix}}), \nonumber
\end{gather} 
where $l_{\text{eff}}$ is the effective spacer length after discounting for allowed mismatches (see Methods). This shows that mismatches exponentially increase the probability of self-targeting, and variable-position mismatches particularly so.
Considering the E. coli system as before, the matching probability increases to $p_{\text{self}} \sim 10^{-4}$ with $k_{\text{fix}} = 5$ nt and $k_{\text{var}} = 5$ nt  (Fig.~\ref{fig:model}b).
Other CRISPR-Cas systems may similarly lie in parameter regimes with appreciable $p_{\text{self}}$,
especially when including indirect self-targeting through priming \cite{nicholson_bioinformatic_2019}.

In the preceding calculations we approximated host and phage genomes as random sequences, neglecting sequence biases that increase or decrease the self-targeting probability depending on whether host and phage genomes share or avoid similar sequence content, respectively. To ascertain the direction and size of this effect, we analyzed genomic data from three host species with the greatest number of associated phage genomes (Appendix \ref{sec:seqcorr}). We found that the self-targeting probability increases modestly (in line with the use of shared oligonucleotide biases for bioinformatic viral-host prediction \cite{Ahlgren2017}), suggesting that the risk of heterologous autoimmunity is actually somewhat larger than predicted by our random model.
All told, our statistical modeling suggests that CRISPR-Cas systems engender an appreciable risk of heterologous autoimmunity.

\begin{figure*}[t]
    \includegraphics[width=\linewidth]{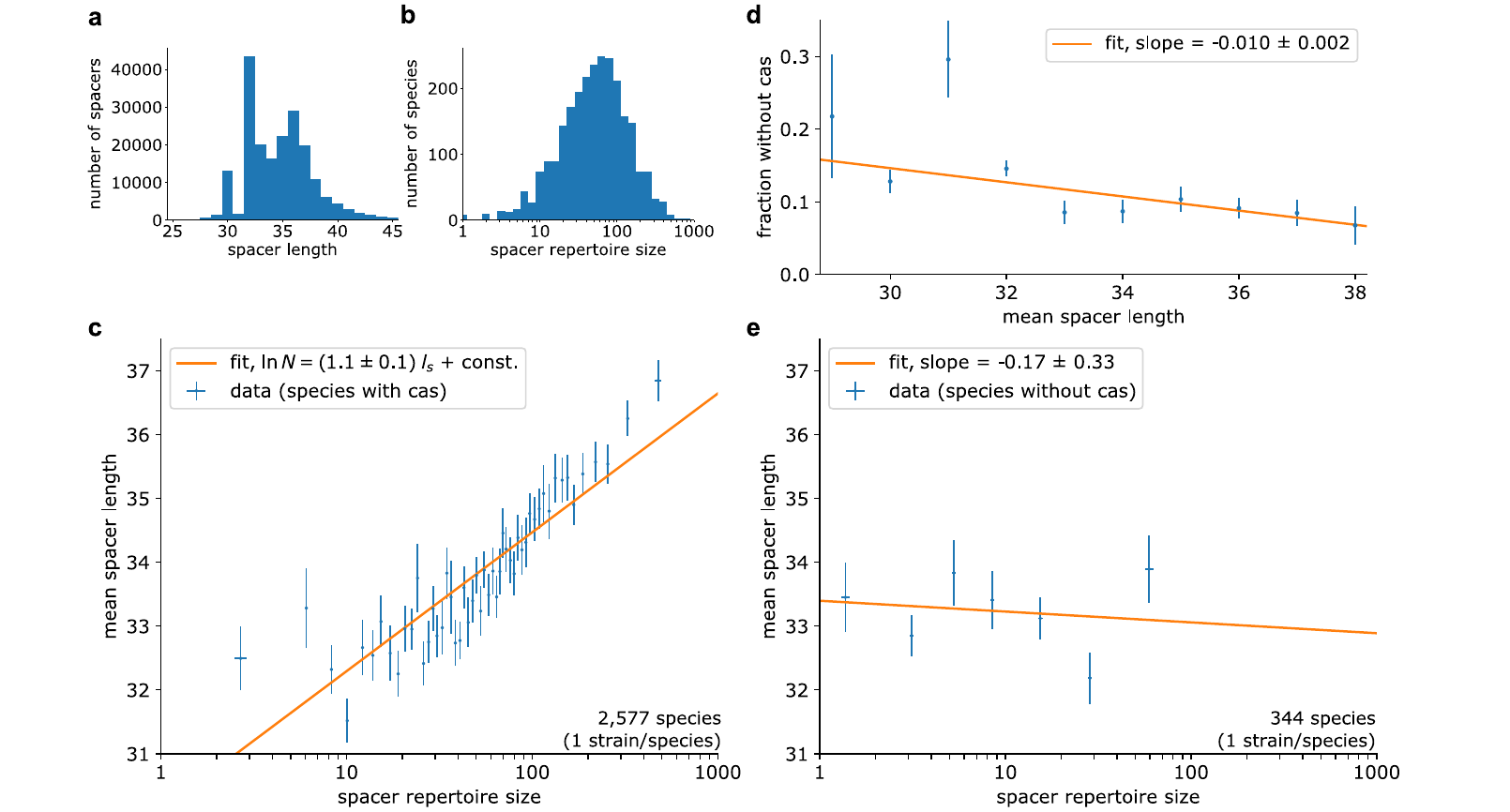}
    \caption{{\bf A scaling law relates CRISPR repertoire size and spacer length.}
    {\bf a}, Distribution of spacer lengths across 2,577 prokaryote strains from unique species carrying CRISPR-cas loci with non-redundant repertoires (see Methods).  {\bf b}, Distribution of repertoire sizes (total number of spacers across CRISPR arrays) for the same strains. Each decade is divided into 10 equal bins in log scale.
    {\bf c}, Mean spacer length increases with log repertoire size, as quantified by a linear fit with positive slope ($p<10^{-6}$, one-sided permutation test). The slope of the fitted regression line matches theory predictions (Eq.~\ref{eq:linear-relation}). Data are shown binned for visualization (50 species/bin, 27 species in rightmost bin), but the fit was to all datapoints (see Fig.~\ref{fig:fit_alldata}). Error bars show standard error of the mean (SEM).
    {\bf d}, The fraction of species missing cas loci decreases with spacer length, as quantified by a linear fit with negative slope ($p = 4 \times 10^{-4}$, one-sided permutation test). 2,577 strains with and 344 strains without cas loci were grouped by mean spacer length, and the fraction of strains without cas was plotted (for all spacer lengths with at least 5 species with and without cas). Error bars show SEM under a binomial assumption.
    {\bf e}, Spacer length and log repertoire size are not associated in strains with nonfunctional CRISPR, as quantified by a linear fit with slope indistinguishable from zero ($p = 0.7$, one-sided permutation test).
    344 strains without cas from unique species and filtered for repertoire non-redundancy were analyzed. Data are shown binned for visualization (50 species/bin, 44 in rightmost bin), but the fit was to all datapoints (see Fig.~\ref{fig:fit_alldata_nocas}).  
    }
    \label{fig:data}
\end{figure*}

\subsection{Mean spacer length scales with repertoire size}

If heterologous autoimmunity is a threat, then spacer length variations, associated with exponential changes in  self-targeting probability (Eqs.~\ref{eq:p_self}--\ref{eq:l_eff}),
should lead to large differences in the cost of maintaining broad repertoires. 
Namely, strains with shorter spacers should have smaller repertoires on average (Fig. \ref{fig:model}c).
A link between receptor binding region size and repertoire breadth has previously been proposed for vertebrate adaptive immunity \cite{Percus1993}, but not CRISPR immunity.
We thus investigated whether spacer length and repertoire size co-vary across prokaryotes.

A tradeoff balancing the benefits of immune defense and the risk of autoimmunity predicts a scaling between spacer length and repertoire size (Appendix~\ref{sec:optimality}). Here we present a simplified model giving the same result. Suppose prokaryotes tolerate a maximum probability $P$ of self-targeting, and that CRISPR systems are selected to maximize protection against pathogens subject to this constraint. Repertoires with $N$ spacers incur a self-targeting probability of $\sim N p_{\text{self}}$ (see Methods), and so we expect selection to drive repertoire growth until $N p_{\text{self}} = P$. Then Eq.~\ref{eq:p_self} implies
\begin{equation}
    \ln N = l \ln 4 - \ln \alpha(k_{\text{fix}}, k_{\text{var}},l) - \ln (L/P).
    \label{eq:scaling}
\end{equation}
Linearizing $\alpha$ around typical spacer lengths $l_0$ (see Methods) predicts that spacer length should scale with the logarithm of repertoire size
\begin{equation}
     \ln N \sim l \left(\ln 4 - \frac{{k_{\text{var}}}}{{l_0 - k_{\text{fix}}}} \right) \sim 1.2 \, l,
    \label{eq:linear-relation}
\end{equation}
where we have used $k_{\rm var} \sim k_{\rm fix} \sim 5$ and $l_0\sim 35$. The slope is robust to variations in these parameter values (Fig.~\ref{fig:fit_model}).

\begin{figure*}[t]
    \includegraphics[width=\linewidth]{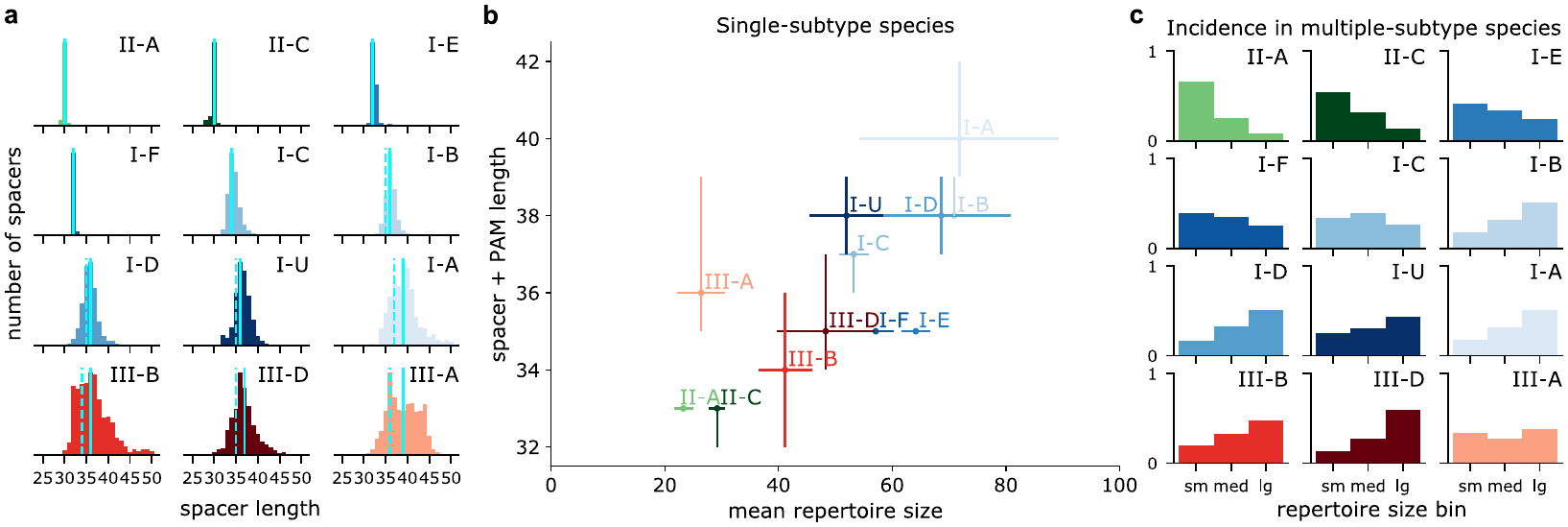}
    \caption{{\bf CRISPR types differ in spacer length preferences and repertoire sizes.} {\bf a}, Spacer length distributions in 1,734 single-type species aggregated by subtype. Subtypes are ordered by
    lower quartile of spacer lengths (dotted lines), with medians also shown (solid lines).
    {\bf b}, Subtypes with larger operative spacer length are associated with larger repertoires (Spearman's $r=0.75$, $p = 3 \times 10^{-3}$, one-sided permutation test). Central dot indicates mean repertoire size and lower quartile of spacer lengths; vertical whisker ranges between lowest decile and median of spacer lengths; and horizontal whisker shows standard deviation of repertoire sizes.
    {\bf c}, In species carrying multiple CRISPR-Cas systems, those with larger repertoires have a lower incidence of types with small spacers. 843 species with multiple cas subtypes were divided into 3 groups of 281 species each having small ($[1, 72]$), medium ($[72, 142]$), and large ($[142, 812]$) repertoires. Bars in subpanels sum to one and indicate relative incidence of subtypes in each group.  The incidence of subtypes with smaller spacers are suppressed in large repertoires, as quantified by the rank correlation of incidence ratios between the large and small bins and operative lengths (Spearman's $r=0.63$, $p = 0.016$, one-sided permutation test). An association also holds when only considering type I subtypes ($r=0.88$, $p = 5 \times 10^{-3}$). For an alternative analysis of incidences across all single- and multiple-subtype species see Fig.~\ref{fig:fit-with-subtypes}.
    }
    \label{fig:castype}
\end{figure*}

We tested this prediction by analyzing a database of CRISPR-Cas systems in publicly available bacterial and archaeal genomes \cite{Pourcel2020,CRISPRCasdb} (see Methods). To sample widely while eliminating oversampling of certain species, we selected strains carrying both CRISPR and cas loci, and picked one strain at random from each species (see Methods). A small fraction of the species shared many spacers with each other, suggesting recent common ancestry of repertoires (Fig.~\ref{fig:filtering}a). To further eliminate such phylogenetic correlations, we clustered species based on shared spacer content and picked one from each cluster (see Methods), producing a non-redundant dataset of 2,577 species sharing zero spacers for further analysis.
We observed a multimodal distribution of spacer lengths acquired by these strains (Fig. \ref{fig:data}a), consistent with narrow spacer length distributions for different CRISPR types (Fig.~\ref{fig:castype}a). The spacer repertoire size distribution, defined as the sum of CRISPR array sizes in each genome, was broad, ranging from 1 to 812 spacers (Fig. \ref{fig:data}b).

Linear regression between spacer length and log-repertoire size of all species with cas gave a slope of $1.0 \pm 0.1$ (Fig. \ref{fig:data}c), in line with the predicted scaling (Eq.~\ref{eq:linear-relation}).
The fit, performed on the 2,577 individual species, is shown in Fig. \ref{fig:data}c with binned data for visualization (all datapoints shown in Fig.~\ref{fig:fit_alldata}).
The empirical law holds over two orders of magnitude in CRISPR repertoire size, but over this range the spacer length changes modestly. These changes however lead to large differences in the self-targeting probability, which is exponential in spacer length (Eq.~\ref{eq:l_eff}).
To assess robustness of the scaling relationship with respect to phylogenetic correlations, we filtered strains on the genus instead of species level and found the same scaling (Fig.~\ref{fig:filtering}b).
A range of cross-reactivity parameters is broadly consistent with this scaling (Fig.~\ref{fig:fit_model}), with a best-fit value of $k_{\text{var}} = 3.82 \pm 0.02$ obtained assuming $k_{\text{fix}} = l_s/6$ (consistent with a 6-nt periodicity in tolerated fixed-position mismatches as in type I-E systems) (Fig.~\ref{fig:fit_model}).
In contrast, we find no association between spacer length and genome length, a potential confounding factor (Fig.~\ref{fig:fit_genomelen}).

Prokaryotes with defective cas genes \cite{Stern2010} or anti-CRISPRs \cite{Watters2018} may tolerate self-targeting. We hypothesized that higher autoimmune risk in species with shorter spacers would lead to a higher rate of cas gene loss, and investigated the incidence of missing cas genes across CRISPR systems. Analyzing another 344 species without cas in the same database \cite{Pourcel2020,CRISPRCasdb} filtered for repertoire non-redundancy, we find that, indeed, a greater fraction of strains with shorter spaces were missing cas (Fig.~\ref{fig:data}d, see Methods).
Once immunopathology from self-targeting is avoided by loss of cas interference genes, spacer length and repertoire size should no longer be related, which we confirmed through a linear fit of all 344 species without cas (Figs.~\ref{fig:data}e and \ref{fig:fit_alldata_nocas}).
These observations strengthen the interpretation of the scaling law as arising from modulation of autoimmune risk by spacer length.

\subsection{Spacer length and repertoire size are associated across CRISPR types}

CRISPR-Cas systems are classified into six types and more than 30 subtypes based on phylogeny and use of different cas genes \cite{Makarova2019}. CRISPR types differ in biophysical mechanisms for spacer acquisition, RNA expression, and interference \cite{Hille2018}, but in all systems base pairing between spacer RNA and target sequence determine specificity. This suggests that a comparison across CRISPR types is possible if we  account for the mechanistic differences between them. We thus defined subtype-specific operative lengths $l$ accounting for subtype-dependent features such as spacer length variability and PAM usage, and studied the association between $l$ and repertoire size across subtypes.

We first categorized the 2,577 species with CRISPR and cas loci by cas subtype, separating species carrying multiple subtypes into a separate group, and focused on subtypes with $>10$ species in the dataset (see Methods and Fig.~\ref{fig:castype}). We quantified the spacer length distributions for the 1,734 species carrying a single subtype (Fig.~\ref{fig:castype}a), and found:
(a) Type II-A and II-C systems have narrow distributions tightly clustered around 30 nt; (b) Type I-E and I-F systems have narrow distributions clustered around 32 nt, while I-A, I-B, I-C, I-D, and I-U have longer and more broadly distributed spacers; (c) Type III-A, III-B, and III-D systems have even longer and more broadly distributed spacers, with median lengths in the 36--39 nt range.
Broader length distributions incur higher autoimmune risk than narrow distributions with the same mean, since the self-targeting probability increases exponentially with decreasing spacer length (Eq.~\ref{eq:l_eff}). We accounted for this by using the first quartile of spacer lengths (Fig.~\ref{fig:castype}a, dashed vertical lines) as a proxy for autoimmune risk where it differed from the median (solid vertical lines).
We further accounted for differences in interference mechanisms between type I and II systems, which require PAM recognition, and type III systems, which do not, by adding a PAM length of 3 nt for types I and II to obtain the subtype-dependent length. Taken together, this definition of the operative length $l$ puts types with and without PAMs, and with broad and narrow length distributions, on a common basis.

Our theory predicts that spacer length should increase with repertoire size, and, as predicted, the length $l$ was positively correlated with mean repertoire size across subtypes (Fig.~\ref{fig:castype}b).
Type II systems have the shortest spacers and additionally process spacer RNA to generate guide RNAs even shorter than the spacer length \cite{Deltcheva2011}. Consistently with our hypothesis, they have the smallest repertoires. Conversely, type I systems have longer spacers and larger repertoires. Comparing between subtypes of type I, shorter spacers are also associated with smaller repertoires. Finally, type III systems have operative lengths $l$ intermediate between type II and type I systems owing to the absence of PAMs and broad spacer length distributions. Correspondingly, their repertoires are of intermediate size.
Given the substantial molecular differences between types, it is striking that a simple biophysically-motivated definition of $l$ largely explains variations in repertoire size.

To further test the predicted association, we analyzed subtype usage preferences among the 843 species carrying multiple subtypes. We divided the data into small, medium and large repertoire size bins of equal size, and determined the relative incidence of subtypes in each bin. We expected that higher autoimmune risk of a subtype (smaller $l$) would lead to lower incidence in large repertoires, and vice versa.
We indeed found a significant positive association between operative length and the incidence ratio between large and small repertoire size bins (Fig.~\ref{fig:castype}c), both globally and among type I subtypes (of which there were a sufficient number to test an association). 
Furthermore, a direct analysis of spacer repertoires among multiple-subtype species shows that a greater proportion of longer spacers is present in larger repertoires (Fig.~\ref{fig:usage_l_m}).

In summary, these analyses show that spacer length distributions of CRISPR subtypes predict repertoire sizes. This is consistent with the hypothesis that the use of different CRISPR-Cas types is shaped by avoidance of heterologous autoimmunity.

\begin{figure*}[t]
    \includegraphics[width=\linewidth]{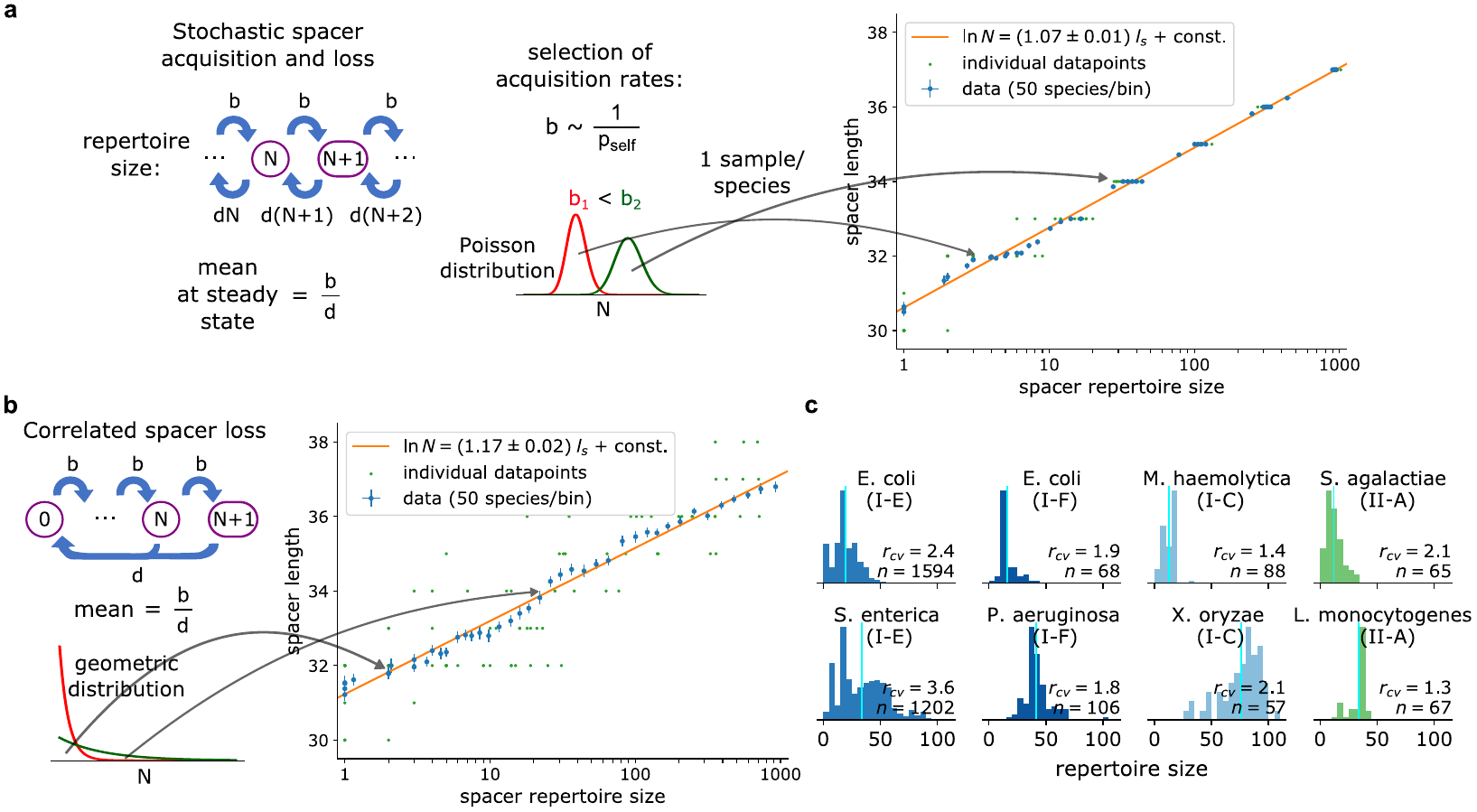}
    \caption{{\bf Repertoire size variability in a stochastic model of spacer acquisition and loss.
    } {\bf a}, 
    Spacers are acquired at a rate $b$ and lost independently at a per-spacer rate $d$ (left). At steady state, repertoire sizes are Poisson distributed with mean $b/d$ (middle panel, see Appendix B), where we have assumed that $b$ has been selected to be inversely proportional to the risk $p_{\text{self}}$ of heterologous autoimmunity. We generated  synthetic data by sampling 2,750 strains from steady-state distributions at different spacer lengths (see Methods). The data show scaling of the mean and variability
    on the single-strain level (right panel). Green points show 100 individually sampled strains; Blue points show means after binning by repertoire size (50 species/bin); Orange is fit to all strains.
    {\bf b}, Correlated spacer loss broadens repertoire size distributions. We solved a model where all spacers are lost simultaneously during a recombination event (left), leading to a geometric steady-state distribution with mean $b/d$ and substantial variability (right, see Appendix B).
    {\bf c}, Prokaryotic strains display large, non-Poissonian variability in repertoire sizes.
    Here, 4 pairs of highly sampled species with identical cas subtypes are displayed. Vertical lines denote species mean. $r_{cv}$ is the ratio of the observed coefficient of variation (CV) to the CV for a Poisson distribution with the same mean. Additionally, there are significant differences in mean repertoire size between species carrying the same subtype ($p < 10^{-6}$ for all 4 pairs, one-sided permutation test).
    }
    \label{fig:evolution}
\end{figure*}

\subsection{Dynamical origin of within-species variability}

While spacer length and repertoire size scale with each other on average, individual strains show substantial variability around this mean (Fig.~\ref{fig:fit_alldata}). This variability 
might arise from other
eco-evolutionary processes shaping repertoire size, such as differences in selective pressures in environments differing in phage diversity, and feedbacks from co-evolving phage
\cite{He2010,Levin2010,Childs2012,Levin2013,Iranzo2013a,Weinberger2012a,Martynov2017,Bradde2017,Han2017}.
Here, in a minimal stochastic model of repertoire size dynamics in a microbial population (Fig.~\ref{fig:evolution}) we show that stochasticity in spacer acquisition and loss suffices to explain a substantial part of this variability.

In our model, spacers are acquired at an effective rate $b$, reflecting both acquisition probability and the purging of lineages that acquire self-targeting spacers \cite{Vercoe2013} (Fig.~\ref{fig:evolution}). For spacer loss the simplest assumption is a constant rate $d$ per spacer (Fig.~\ref{fig:evolution}a), so that the steady-state repertoire size obeys a Poisson distribution with mean $b/d$ (Appendix~\ref{sec:dy}). Assuming that a selection principle drives $b$ to be inversely proportional to the self-targeting probability (as in Sec. B), this model predicts a scaling law between spacer length and the mean repertoire size as in Eq. \ref{eq:scaling}.
We generated a synthetic dataset by sampling strains from steady-state distributions with different $l$ and hence mean repertoire size (Fig.~\ref{fig:evolution}a, middle; see Methods), and found that the predicted scaling was obeyed (Fig.~\ref{fig:evolution}a, right),
but Poisson variation around this relation is insufficient to explain the variability in the actual data, especially for large repertoires (Fig.~\ref{fig:fit_alldata}).  However, spacer loss occurs through double-stranded DNA breaks followed by homologous recombination at a different CRISPR repeat, which deletes chunks of an array in a single deletion event (see e.g. \cite{Tyson2008,Horvath2008}). Such correlated spacer loss greatly increases variability. We considered a simple limiting model of correlated spacer loss where entire arrays are deleted at once, finding analytically that repertoire sizes obey a much broader geometric distribution at steady state (Appendix~\ref{sec:dy}). A synthetic dataset generated from the revised model shows broad variability of repertoire sizes around the mean relationship that persists even for large repertoires (Fig. \ref{fig:evolution}b).

To quantify the variability in real microbial populations using the same CRISPR-Cas system, we focused on four pairs of highly-sampled species carrying the same CRISPR-Cas subtype (see Methods, Fig.~\ref{fig:evolution}c). We observed broad, non-Poissonian repertoire size distributions, with  variability within species comprising a substantial part of the overall variance, consistently with effects  induced by stochastic spacer acquisition and loss.
Additional variability,  leading to different mean repertoire sizes across  species with the same CRISPR-Cas subtype, may originate from different microbes inhabiting environments that differ in viral diversity and thus pressure to acquire broad immune defense. We modeled the effect of between-species variability by sampling individuals from steady-state distributions where the prefactor in $b \propto 1/p_{\text{self}}$ is additionally drawn from a wide distribution (see Methods).  Including this  further increases the variations between individually sampled strains (Fig.~\ref{fig:simulated_Alognormal}), resembling the high  variability observed across prokaryotes (Fig.~\ref{fig:fit_alldata}) while still preserving the scaling of the means.

\section{Discussion}

An adaptive immune system is dangerous equipment to own: immune receptors, intended as defenses against foreign invaders, risk targeting the self instead. In CRISPR-Cas systems, biophysical mechanisms avoiding various forms of autoimmunity such as  CRISPR locus targeting and self-spacer acquisition are known \cite{Marraffini2015,Deveau2008,Semenova2011,Levy2015}. Here, we propose that heterologous autoimmunity, where spacers acquired from foreign DNA seed self-targeting, are an additional and potentially significant threat to microbes carrying CRISPR-Cas.
This threat is analogous to off-target effects in genome-editing \cite{Fu2013,Hsu2013}, and has been observed experimentally \cite{Staals2016}, although wider implications for the evolution and diversity of CRISPR systems are unexplored.
We showed that avoidance of this form of autoimmunity while maximizing antiviral defense predicts a scaling law relating spacer length and CRISPR repertoire size. The scaling depends on the number and nature of sequence mismatches permitted during CRISPR interference and primed acquisition.

To test our prediction, we performed a comparative analysis exploiting natural variation in CRISPR-Cas systems across bacterial and archaeal species, and demonstrated that: (a) the predicted scaling law is realized, (b) the observed scaling constrains parameters for cross-reactive CRISPR targeting to lie in a range consistent with experimental studies, (c) the scaling arises in part from differential use of different CRISPR-cas subtypes with different spacer length distributions, and (d) variations of individual strains around the scaling law arise partly from stochastic spacer acquisition and loss.  In addition, we demonstrated a negative control: CRISPR arrays in species that no longer have functional Cas proteins, and thus are not at risk of autoimmunity, do not show the predicted scaling relation.

Our statistical theory requires that effective acquisition rates should decrease as spacer length decreases. There are two mechanisms by which selection could lead to such a dependence. First, the negative fitness effect of self-targeting \cite{Vercoe2013} purges lineages that undergo deleterious acquisition events. Indeed, CRISPR arrays are selected for the absence of self-targeting spacers \cite{Stern2010}. Effectively, this reduces the net acquisition rate among surviving lineages. Second, over longer evolutionary timescales, different CRISPR-Cas systems may be selected to acquire spacers at different rates depending on their respective risks of autoimmunity. These differences in rates could arise from the maintenance of multiple copies of cas genes, or through regulation of cas expression \cite{patterson_regulation_2017}. Indeed, spacer repertoire size increases with the number of cas loci (Fig. \ref{fig:numloci}),  suggesting that larger gene copy numbers of cas1 and cas2, necessary for spacer acquisition, result in greater acquisition rates. Interestingly, strains having exactly one copy of both cas1 and cas2 still obey a scaling relationship (Fig. \ref{fig:fit_1copycas}), suggesting that regulation of these genes also contributes to minimizing autoimmune risk \cite{patterson_regulation_2017}.

We propose two further tests of the link between spacer length and autoimmune risk:
(1) A bioinformatic demonstration that self-targeting spacers several mismatches away (going beyond exact matches studied in the literature \cite{Stern2010,Watters2018,Nobrega2020}) are associated with non-canonical PAMs in the protospacer, mutations in the adjoining repeat or cas locus, the spacer being closest to the leader end of the array, and/or  increased incidence of anti-CRISPR genes \cite{Stern2010,Watters2018,Nobrega2020}; and
(2) An experimental demonstration that CRISPR-Cas subtypes with shorter spacers (types II-A and II-C for example) acquire spacers at a slower rate than those with longer spacers (for example types I-A, I-B and type III).

Our comparative analysis is enabled by a recent explosion in the number of sequenced microbial genomes and the discovery of considerable variation in CRISPR-Cas systems. We have used this data to sample diverse species, and  to filter out CRISPR repertoires that contain overlapping spacers which might be related by recent horizontal transfer or shared ancestry of CRISPR-cas loci.
The resulting dataset can be thought of as independent samples of spacer repertoires created {\it de novo} as a result of distinct environments, viral threats, and selective pressures faced by the respective species. We also find that the scaling relationship is robust to filtering at the genus instead of the species level (Fig.~\ref{fig:filtering}b), which  removes essentially all spacer overlaps between repertoires.

An alternative hypothesis that may explain the scaling law relies on the {\it beneficial} effects of cross-reactivity.  Shorter, cross-reactive spacers enable broader immunity and face a lower threat of mutational escape by viruses than longer spacers; thus, fewer of them are needed to achieve the same breadth of immunity. This alternative hypothesis alone, however, does not predict selection pressure for losing cas  that is associated to spacer length  (Fig.~\ref{fig:data}d).  We expect models explicitly balancing costs of autoimmunity with benefits of broad defense to show a similar scaling law (a simple example appears in Appendix A). A comprehensive analysis of such models is an important direction for future work, but will likely require more detailed knowledge of parameters such as the frequency of viral infection and viral mutation rate in natural environments.

A similar tradeoff between sensitivity to pathogens and autoimmune risk shapes the evolution of vertebrate adaptive immune systems \cite{Percus1993,Metcalf2017}. In light of our results, it would be interesting to determine whether this tradeoff also leads to a relation between the size of the immune repertoire and immune specificity in vertebrates. Such a relation will likely be harder to ascertain for vertebrates as patterns of cross-reactivity between lymphocyte receptors and antigens are more complex. Interestingly, however, T cell receptor hypervariable regions in human are several nucleotides longer on average than those found in mice \cite{Sethna2017}, which accompanies a substantial increase in repertoire size in human. If longer hypervariable regions translate to a greater specificity on average, one might view the increased human receptor length as an adaptation to a larger repertoire. The key to our current work was the ability to compare microbial immune strategies across a large panel of phylogenetically distant species. Further insight into how this tradeoff shapes vertebrate immune systems might thus be gained  by building on recent efforts to survey adaptive immune diversity in a broader range of vertebrates \cite{Castro2017,Covacu2016}.

Many theoretical studies of adaptive immunity in both prokaryotes \cite{He2010,Levin2010,Childs2012,Levin2013,Iranzo2013a,Weinberger2012a,Martynov2017,Bradde2017,Han2017} and vertebrates \cite{Antia2005,Lythe2016,Desponds2017,Gaimann2020} consider detailed dynamical models of evolving immune repertoires. For prokaryotes, such dynamical models can be regarded as describing the role of CRISPR-Cas as a short-term memory for defense against a co-evolving phage \cite{Bradde2020}.  Studying adaptive immunity in this way requires detailed knowledge of the parameters controlling the dynamics, many of which are not well-characterized experimentally. In this paper, we took an alternative approach of focusing on the statistical logic of adaptive immunity, where we regard the bacterial immune system as a functional mechanism for maintaining a long-term memory \cite{Bradde2020} of a diverse phage landscape \cite{edwards2005viral}, via probabilistic matching of genomic sequences.  Previous work taking this perspective offered an explanation for why prokaryotic spacer repertoires lie in the range of a few dozen to a few hundred spacers \cite{Bradde2020}. As in our discussion of possible  mechanisms for generating the observed scaling law, evolution should select dynamics that achieve the statistical organization that we predict, because this is what is useful for achieving a broad defense against phage while avoiding autoimmunity. A probability theory perspective of this kind has been applied to the logic of the adaptive immune repertoire of vertebrates \cite{Mayer:2015,Mayer2016,mayer2019well},
but to our knowledge we are presenting a novel approach to the study of CRISPR-based autoimmunity.

\section{Materials and Methods}

\paragraph{Derivation of self-targeting probability.}

We estimate the probability of an alignment between a spacer + PAM sequence of length $l$ and a host genome of length $L$. We assume that both sequences are random and uncorrelated, with nucleotide usage frequencies of 1/4. In a length-$L$ genome, where $L \gg l$, there are $L - l + 1 \approx L$ starting positions for an alignment. The matching probability at each position, $p_m$, depends on the number and nature of mismatches tolerated. In regimes where $p_m$ is small, the matching probabilities at the different positions may be treated independently. Thus, the probability of having at least one alignment within the length-$L$ genome is
\begin{align}
    p_{\text{self}} &= 1 - (1 - p_m)^{L-l+1} \nonumber \\
    &\approx L p_m, \text{ since } p_m \ll 1, l \ll L.
\end{align}

If no mismatches are tolerated, $p_m = 4^{-l}$ as in Eq.~\ref{eq:p_0}.
At each site where a mismatch is allowed, four alternative nucleotide choices are possible. This gives a number $\alpha$ of unique complementary sequences matching a given spacer, which we compute as a function of the number and nature of mismatches.
If up to $k_{\text{fix}}$ mismatches are tolerated at fixed positions in the alignment, $\alpha = 4^{k_{\text{fix}}}$. If instead up to $k_{\text{var}}$ mismatches are tolerated anywhere in the complementary region, naively $\alpha \sim \binom{l}{k_{\text{var}}} 4^{k_{\text{var}}}$, where the binomial coefficient is the number of combinations of sites where mismatches are allowed. This however overcounts matching sequences, and the precise expression is
\begin{align}
    \alpha = \sum_{i=0}^{k_{\text{var}}} \binom{l}{i} 3^i,
\end{align}
where each term in the sum is the number of unique complementary sequences having exactly $i$ mismatches. The largest term dominates, giving $\alpha \approx \binom{l}{k_{\text{var}}} 3^{k_{\text{var}}}$.
Thus, combining $k_{\text{fix}}$ mismatches at fixed positions and up to $k_{\text{var}}$ mismatches at any of the remaining $l - k_{\text{fix}}$ positions gives
\begin{equation}
    \alpha(k_{\text{fix}},k_{\text{var}},l) \approx 4^{k_{\text{fix}}} \binom{l - k_{\text{fix}}}{k_{\text{var}}} 3^{k_{\text{var}}}.
    \label{eq:alpha}
\end{equation}
To guide intuition we introduce an effective spacer length, $l_\text{eff}$, by $p_m \equiv  4^{-l_{{\rm eff}}}$. To leading order the binomial expression in Eq.~\ref{eq:alpha} is approximated by $(l-k_{\text{fix}})^{k_{\text{var}}}$. This gives $l_{{\rm eff}} \approx l - k_{{\rm fix}} - k_{{\rm var}} \log_4 3(l - k_{{\rm fix}})$ as in Eq.~\ref{eq:l_eff}.

The probability that a repertoire of $N$ spacers avoids self-targeting, $1 - P_\text{self}$, is one minus the probability that at least one spacer targets self. This gives
\begin{align}
    P_\text{self} &= 1 - (1 - p_\text{self})^N\nonumber \\
    &\approx N p_\text{self}, \text{ since } L p_m \ll 1.
\end{align}
If CRISPR repertoires are selected to maximize repertoire size subject to the constraint $P_\text{self} \leq P$, we obtain Eq.~\ref{eq:scaling}. Taylor expanding $\ln N$ around $l = l_0$ gives Eq.~\ref{eq:linear-relation} to lowest order in $l$.

\paragraph{Comparative analyses.}

To perform our comparative analyses we used data from CRISPRCasdb \cite{Pourcel2020}, a database of CRISPR arrays and cas loci identified in public bacterial and archaeal whole-genome assemblies using CRISPRCasFinder \cite{Couvin2018,CRISPRCasdb}. Predicted CRISPR arrays are assigned evidence levels 1--4, 4 being the highest confidence \cite{Couvin2018}. We restricted our analyses to level 4 CRISPR arrays only. Strains carrying both CRISPR arrays and cas loci were used for the analyses producing Figs. \ref{fig:data}a--d, \ref{fig:castype}a--c, and \ref{fig:evolution}c. Strains carrying CRISPR arrays but no cas loci were used for the analyses producing Figs. \ref{fig:data}d--e. Before performing these analyses, we eliminated oversampling of certain species in the database by picking one strain at random per species (2,730 species with cas, and 369 species without cas). To further eliminate phylogenetic correlations between repertoires, some of which contain many shared spacers (Fig.~\ref{fig:filtering}a), we performed single-linkage clustering of these species based on shared spacers, and picked one species from each cluster at random, producing a non-redundant dataset of repertoires sharing zero spacers between them for further analysis (2,577 species with cas, and 344 species without cas).

To produce Fig.~\ref{fig:castype}a--c, the randomly chosen species were classified by annotated cas subtype, or into a separate group if they carry multiple subtypes. The 12 subtypes that have $>\! 10$ species represented in CRISPRCasdb are shown in Fig.~\ref{fig:castype}. Among the 4 subtypes having at least 2 species with $>\! 50$ sequenced strains in CRISPRCasdb (types I-E, I-F, I-C, and II-A), we displayed the 2 most highly sampled species for each subtype in Fig.~\ref{fig:evolution}c.

\paragraph{Statistical significance tests.}

In Figs.~\ref{fig:data}c,e significance of slope was assessed by shuffling repertoire size values among strains and recomputing the slope. In Fig. \ref{fig:data}d significance of slope was assessed by shuffling the assignment of cas or no cas at each spacer length and recomputing the slope. In Figs.~\ref{fig:castype}b--c significance of association was assessed by shuffling repertoire size and incidence ratio values, respectively, among subtypes and recomputing Spearman's $r$. In Fig.~\ref{fig:evolution}c significance of difference in mean was assessed by shuffling the assignment of species to each repertoire size (for each pair of species carrying the same CRISPR-Cas subtype) and recomputing the means between the 2 resulting populations. In all cases the p-value is the number of times shuffling produced a value (slope, Spearman's $r$, or difference in mean) greater than the one observed in $10^6$ shuffles (i.e., one-sided permutation test).

\paragraph{Synthetic data generation and analysis.}

A synthetic dataset producing a scaling law was generated in the following way:
(1) A spacer of length $l_s$ was drawn from the length distribution of Fig.~\ref{fig:data}a, and
(2) a repertoire size distribution with mean $A/p_\text{self}$ was created, from which one strain was sampled and added to the dataset. Parameter values of $L = 5 \times 10^6$, $l_p = 3$, $k_\text{fix} = l_s/6$, $k_\text{var} = 3$, and $A = 10^{-5.5}$ were used. The steady-state distributions are Poisson in Fig.~\ref{fig:evolution}a, and geometric with the same mean in Fig.~\ref{fig:evolution}b.

{\bf Acknowledgements.}
We thank Serena Bradde and Madeleine Bonsma-Fisher for helpful comments on this paper.
VB and HC were supported in part by a Simons Foundation grant in Mathematical Modeling for Living Systems (400425) for Adaptive Molecular Sensing in the Olfactory and Immune Systems, and by the NSF Center for the Physics of Biological Function (PHY-1734030). AM was supported by a Lewis–Sigler fellowship. VB thanks the Aspen Center for Physics, which is supported by NSF grant PHY-160761, and the Santa Fe Institute for hospitality during this work.

{\bf Significance.}
CRISPR has become a major tool for genome-editing in biotechnology and medicine. A serious concern for this application is the prevalence of off-target effects, where edits are made to unintended genes because of cross-reactive targeting. Here we show that such effects are also a constraint in CRISPR’s natural role as an adaptive immune system in prokaryotes. Our work highlights a fitness tradeoff between breadth of protection and the risk of autoimmunity that generates a scaling law in the organization of immunity across species.

\bibliography{library}

\clearpage
\appendix

\section{An explicit model for optimal repertoire size}\label{sec:optimality}

CRISPR repertoires evolve via spacer acquisition and loss. New spacers that target viruses increase protection of the host from threats, whereas self-targeting spacers cause autoimmunity. Here we construct a simple model of this tradeoff, and find the repertoire size $N$ that maximizes long-term population growth rate given the competing constraints of antiviral immunity and avoidance of heterologous autoimmunity.

We consider a colony of bacteria or archaea that are attacked episodically by a random virus from a diverse global pool. 
Following each attack many individuals in the colony will die, but so long as some survive the colony will be replenished. In the regime of large viral diversity, there is a low probability $P_{\text{def}}(N)$ that a given spacer in a lineage is specific to a re-emerging threat, but larger repertoires increase this probability. The precise dependence of the probability of immune defense, $P_{\text{def}}(N)$, on repertoire size involves detailed epidemiological characteristics of host-phage co-evolution, but in general we expect a sublinear dependence because of diminishing returns \cite{Bradde2020}. Here, for simplicity we assume the general functional form
\begin{equation}
    P_{\text{def}}(N) = (N/\beta)^\gamma,
\end{equation}
where $\beta$ and $\gamma < 1$ are fixed parameters.

A larger repertoire incurs a greater risk of autoimmunity. Assuming autoimmunity is fatal, survival requires that none of the $N$ spacers are self-targeting. The probability $1-P_{\text{auto}}(N)$ of avoiding autoimmunity is thus
\begin{equation}
     1 - P_{\text{auto}}(N) = (1-p_{\text{self}})^N \approx e^{-N p_{\text{self}}}. 
\end{equation}

To find the optimal repertoire size, we need to determine how $P_{\text{def}}$ and $P_{\text{auto}}$ combine to impact population fitness. To this end, we make use of a discrete-time model of population growth in a randomly fluctuating environment in its large population size limit \cite{Mayer2017b}. We assume that viral epidemics occur at probability $q$ per generation. The expected number of offspring $f_t$ is reduced from its maximal value $e^{R_{max}}$ to
\begin{equation}
    f_t = e^{R_{max}} P_{\text{def}}^{x_t} (1 - P_{\text{auto}}),
\end{equation}
where $x_t = 1$ in the presence of a viral epidemic and $x_t = 0$ otherwise.
The succession of booms and busts determines the population size at time $T$, $N_T = N_0 \prod_t f_t$, where $N_0$ is the initial population size. The long-term population growth rate, defined as 
\begin{equation}
    \Lambda = \lim_{T\rightarrow \infty} \frac{1}{T} \ln \frac{N_T}{N_0},
\end{equation}
is the evolutionarily relevant measure of long-term success of different strategies \cite{Mayer2017b}.
Applying the law of large numbers, we obtain an expression for the long-term population growth rate for this model,
\begin{equation}
    \Lambda \approx R_{max} + q \gamma \ln \frac{N}{\beta} - N p_\text{self}.
\end{equation}

To find the optimal repertoire size $N$, we differentiate $\Lambda$ with respect to $N$ and set it to 0:
\begin{align}
    &\phantom{\Rightarrow\;\;} \frac{\partial \Lambda}{\partial N} = \frac{q\gamma}{N} 
    - p_{\text{self}} = 0 \nonumber\\
    &\Rightarrow N  = \frac{q \gamma}{p_{\text{self}}} 
    \nonumber\\
    &\Rightarrow \ln N = l_\text{eff} \ln 4 - 
    \ln \frac{L}{q \gamma},
\end{align}
where we have used Eq.~\ref{eq:l_eff} for $p_{\rm self}$.  Substituting $l_{\rm eff} = l - \log_4 \alpha$ we obtain
\begin{equation}
\ln N = l \ln 4 -\ln \alpha - 
\ln \frac{L}{q \gamma},
\label{tradeoffresult}
\end{equation}
since $\log_4\alpha \ln 4 = \ln \alpha$.
We thus recover Eq.~\ref{eq:scaling} of the main text, where $q \gamma$ plays the role of the parameter $P$.

\section{Models of stochastic spacer acquisition and loss}\label{sec:dy}

Consider a host population acquiring spacers of length $l$. Let the number of individuals in the population that have repertoire size $n$ ($n\geq 0$) be $X_n$. Consider spacer acquisition to occur at a rate $b$:
\begin{align}
    X_n \xrightarrow{b} X_{n+1}.
\end{align}
Spacer acquisition is balanced by spacer loss leading to a well-defined steady-state distribution of repertoire size. Spacer loss occurs through double-stranded DNA breaks followed by homologous recombination at a subsequent repeat, which deletes chunks of the CRISPR array (see e.g. \cite{Tyson2008,Horvath2008}). The precise rate and mechanism by which this occurs is not well-understood. Here, we consider 3 solvable scenarios of this process:
\begin{align}
    \text{1: } &X_n \xrightarrow{d} X_{n-1} \\
    \text{2: } &X_n \xrightarrow{dn} X_{n-1} \\
    \text{3: } &X_n \xrightarrow{d} X_{0}.
\end{align}
The first scenario represents spacer loss at the end(s) of the CRISPR array, hence independent of $n$. The second represents a constant per-spacer loss rate. For the third scenario, all spacers are lost in a deletion event, which is a solvable limit of several spacers being deleted at a time.

\paragraph*{Scenario 1: $X_n \xrightarrow{d} X_{n-1}$.}

The probabilities $P_n$ ($n \geq 0$) obey the following master equation:
\begin{align}
    \frac{\ud P_n}{\ud t} &= -(b+d)P_n + bP_{n-1} + dP_{n+1},\ n\geq 1 \label{eq:dy1a}\\
    \frac{\ud P_0}{\ud t} &= -bP_0 + dP_1.\label{eq:dy1b}
\end{align}
The steady state fulfills the detailed balance condition,
\begin{equation}
    dP_n = bP_{n-1}.
    \label{eq:db1}
\end{equation}
We can solve the recursion equation (Eq.~\ref{eq:db1}) for the steady-state distribution,
\begin{align}
    P_n = (b/d)^n (1 - b/d),
    \label{eq:d1}
\end{align}
which is geometric with parameter $1 - b/d$.
Its mean is $b/(b-d)$, implying that a well-defined steady state is only possible if $d > b$.

\paragraph*{Scenario 2: $X_n \xrightarrow{dn} X_{n-1}$.}

The master equation is:
\begin{align}
    \frac{\ud P_n}{\ud t} &= -(b+dn)P_n + bP_{n-1} + d(n+1)P_{n+1},\ n\geq 1 \label{eq:dy2a}\\
    \frac{\ud P_0}{\ud t} &= -bP_0 + dP_1.\label{eq:dy2b}
\end{align}
At steady state again detailed balance holds:
\begin{equation}
    dnP_n = bP_{n-1}.
    \label{eq:db2}
\end{equation}
Eq.~\ref{eq:db2} implies that the steady-state distribution is Poisson with mean $b/d$:
\begin{align}
    P_n = \frac{1}{n!} (b/d)^n e^{-b/d}.
    \label{eq:d2}
\end{align}

\paragraph*{Scenario 3: $X_n \xrightarrow{d} X_0$.}

Here, the master equation is:
\begin{align}
    \frac{\ud P_n}{\ud t} &= -(b+d)P_n + bP_{n-1},\ n\geq 1 \label{eq:dy3a}\\
    \frac{\ud P_0}{\ud t} &= -bP_0 + d (1-P_0).\label{eq:dy3b}
\end{align}
Here there is no detailed balance, but probability flux is conserved:
\begin{equation}
    (d+b)P_n = bP_{n-1}.
    \label{eq:db3}
\end{equation}
Eq.~\ref{eq:db3} implies that the steady-state distribution is geometric with parameter $d/(b+d)$:
\begin{align}
    P_n = \left[\frac{b}{b + d}\right]^n \frac{d}{b + d}.
    \label{eq:d3}
\end{align}
The mean of this distribution is $b/d$.

\section{Sequence correlations and the probability of heterologous autoimmunity} \label{sec:seqcorr}

In the main text, we assumed that phage-derived and host-derived sequences are completely random in order to derive order-of-magnitude estimates for coincidence probabilities.
These probabilities might change if we account for sequence correlations between prokaryotes and the phage that prey on them. To test for this possibility, here we study three prokaryotes and their associated phages to estimate how  correlations change the probability of matching sequences.

\begin{figure}[h]
    \includegraphics{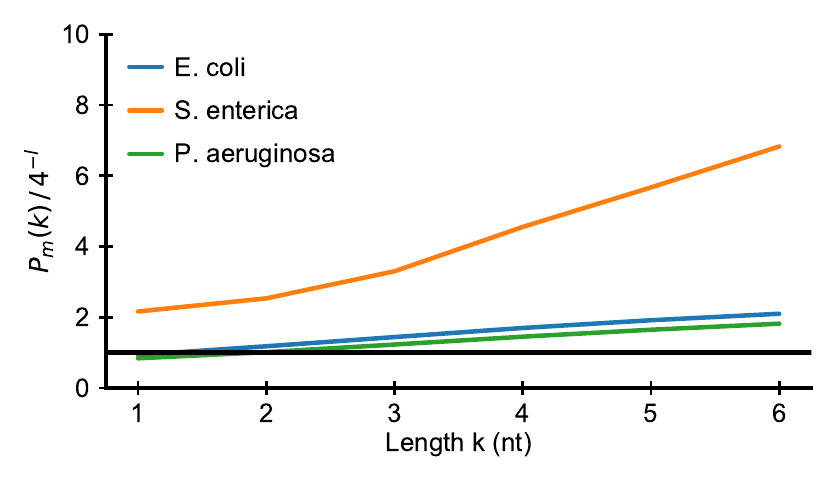}
    \caption{Probability of sequence matching between random phage and host sequences of length $l=35$ constructed from kmers of increasing length (Eq.~\ref{eq:P_kmer}), relative to the random expectation $P_m = 4^{-l}$. Sequence correlations increase coincidence probabilities, such that the risk of heterologous autoimmunity is increased relative to the predictions of the random model.}
    \label{fig:seqcorr}
\end{figure}

We downloaded the genomes of all phages associated with three host species having the greatest number of known phages from public sequencing databases via \url{https://www.ncbi.nlm.nih.gov/labs/virus/vssi/}. The hosts are E. coli (taxon id 562), S. enterica (taxon id 287), and P. aeruginosa (taxon id 28901). To reduce redundancy we clustered the phage dataset at 90\% sequence identity using MMSeqs2 \cite{Steinegger2017}. After this clustering 726 (E. coli), 285 (S. enterica), and 380  (P. aeruginosa) non-redundant phage genomes remained for the three hosts, with a total of $6.6 \cdot 10^7$ nt (E. coli), $1.9 \cdot 10^7$ nt (S. enterica), and $3.1 \cdot 10^7$ nt (P. aeruginosa), respectively. We divided each genome into all overlapping sequences of length $k$ and tabulated each of the $4^k$ kmer frequencies. We then used these to calculate the probability of a match for sequences of length $l$, assuming the full sequence is composed of independent kmers:
\begin{equation}
    P_m(k) = \left(\sum_i p_i q_i\right)^{l/k} \, .
    \label{eq:P_kmer}
\end{equation}
Here the sum runs over all kmers, and $p_i$ and $q_i$ are empirical kmer probabilities in the host and phage genomes, respectively. By varying $k$ we can assess how different sequence features (from nucleotide biases to higher order correlations) influence coincidence probabilities.
In all cases the probability of heterologous autoimmunity increases relative to the baseline expectation, in line with the use of shared oligonucleotide biases for bioinformatic viral-host prediction \cite{Ahlgren2017}.
Deviations from the baseline theory are modest as compared to the exponential dependence on spacer length, as each additional nucleotide in spacer length decreases $P_m$ by a factor of $\sim 4$.

\newpage

\onecolumngrid
\subsection*{SI figures}

\setcounter{figure}{0}
\renewcommand{\thefigure}{S\arabic{figure}}

\begin{figure}[h]
    \includegraphics{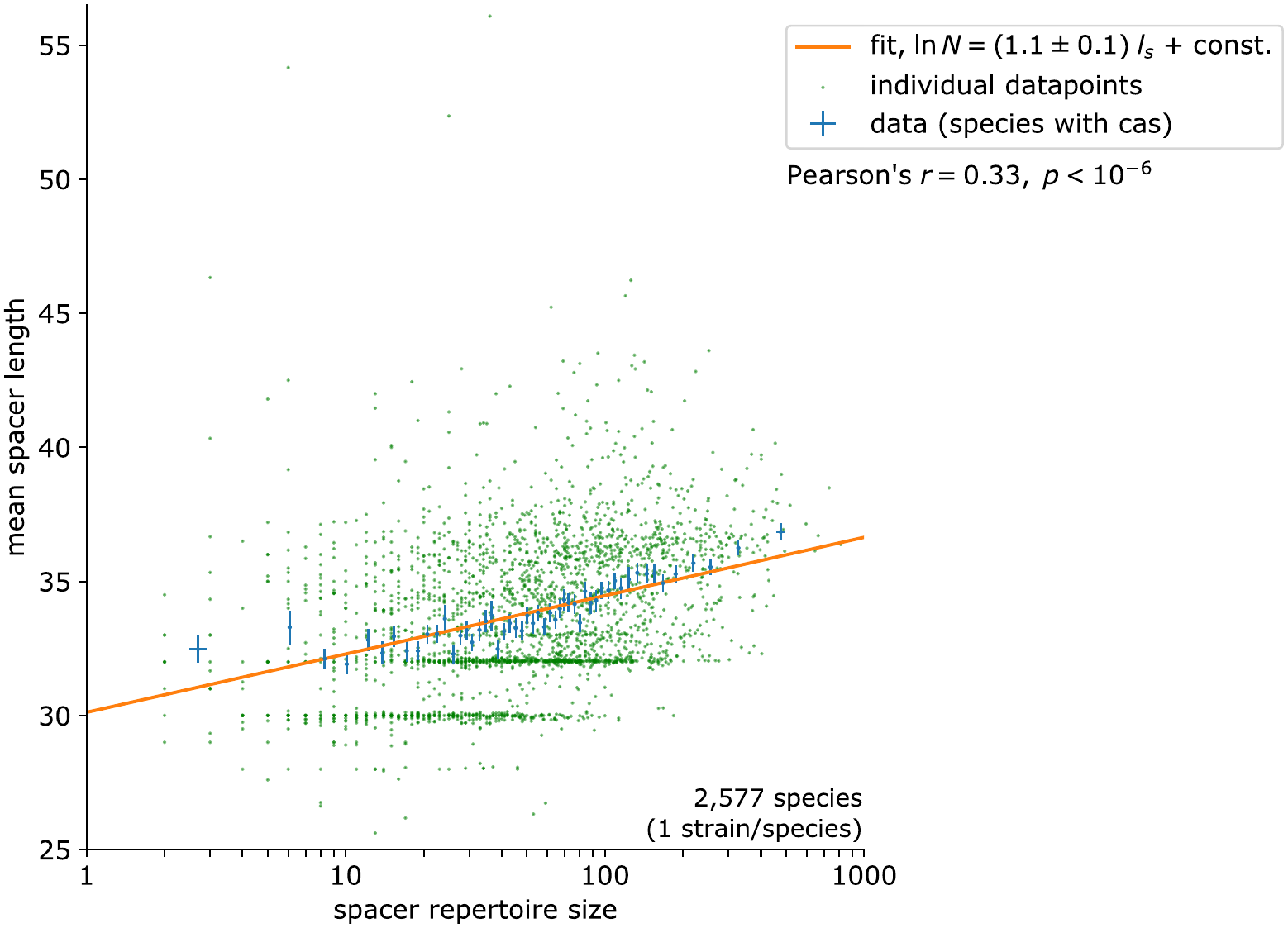}
    \caption{A scaling law between CRISPR repertoire size and mean spacer length in species with cas. The blue points are data from 2,577 species binned in increasing windows of repertoire size (50 species/bin) with error bars denoting the standard error of the mean. The green points are all 2,577 individual datapoints, and the orange line is the linear fit to the individual species as in Fig.~\ref{fig:data}c. Dense horizontal bands along mean spacer lengths 30 nt and 32 nt reflect many species acquiring spacers of these lengths (see Fig.~\ref{fig:data}a). 
    }
    \label{fig:fit_alldata}
\end{figure}

\begin{figure}[h]
    \includegraphics{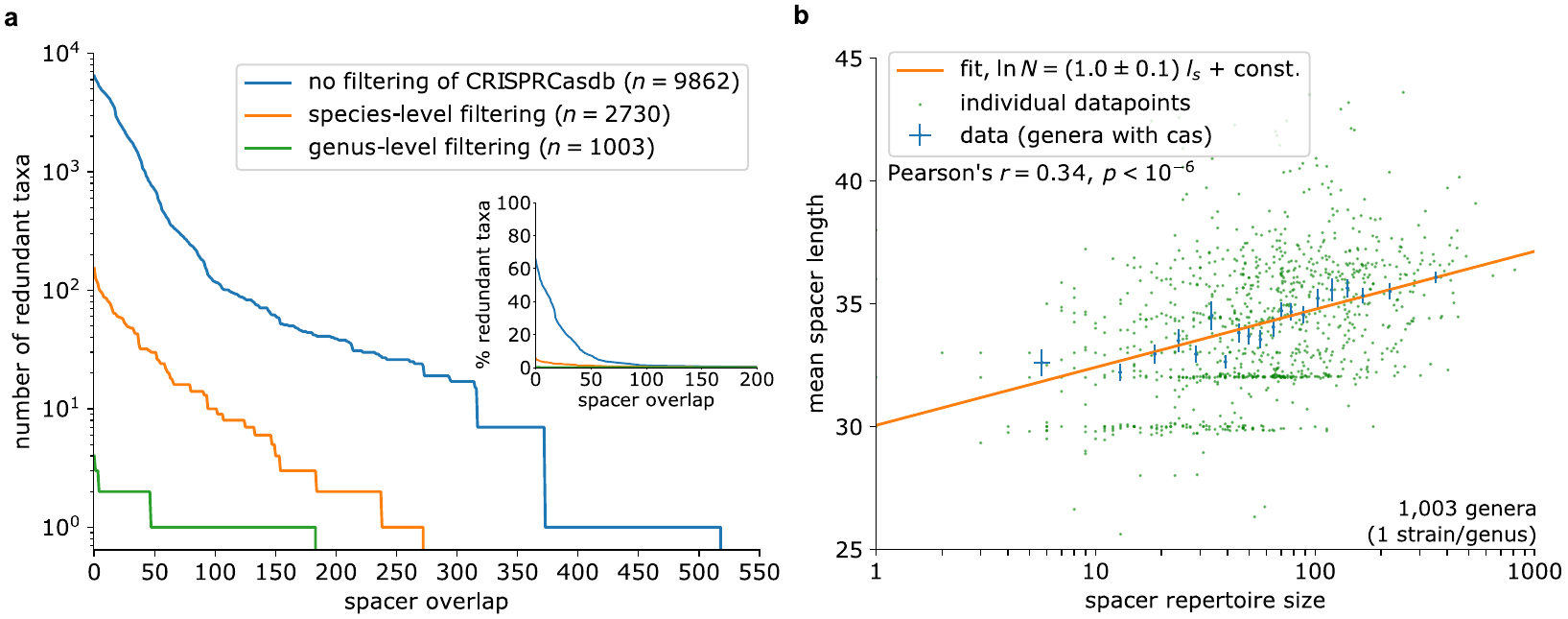}
    \caption{Phylogenetic considerations for a comparative analysis of CRISPR-Cas systems. {\bf a}, High overlap of spacer repertoires between strains carrying CRISPR and cas loci. Single-linkage clustering of spacer repertoires based on the number of shared spacers was performed on all strains carrying CRISPR and cas loci, either without filtering (blue line) or after filtering by species (orange line) or genus (green line). The dendrogram was cut at different thresholds of spacer overlap (x-axis), and the number of redundant strains (i.e., those sharing that many spacers with a non-redundant strain) are plotted. Without filtering of CRISPRCasdb, more than 60\% of strains are redundant at zero spacer overlap (inset), and many strains share dozens or hundreds of spacers with another. A large part of this arises from closely related strains and CRISPR-cas loci. Filtering strains by species, most of this redundancy is removed (inset). The remaining redundancy might arise from shared ancestry between closely related species or recent horizontal transfer of CRISPR-cas loci between species. In the main text, filtering out redundant species was performed such that all spacer repertoires in the dataset used for analysis do not share any spacers. Alternatively, filtering at the genus level (green line) removes essentially all overlaps.
    {\bf b}, The scaling law is robust to different levels of filtering. The blue points are data from 1,003 genera binned in increasing windows of repertoire size (50 genera/bin, 53 genera in the rightmost bin) with error bars denoting the standard error of the mean. The green points are all 1,003 individual datapoints, and the orange line is the linear fit to the individual species. 4 datapoints with mean spacer length $>\! 45$ nt were not displayed.
    }
    \label{fig:filtering}
\end{figure}

\begin{figure}[h]
    \includegraphics{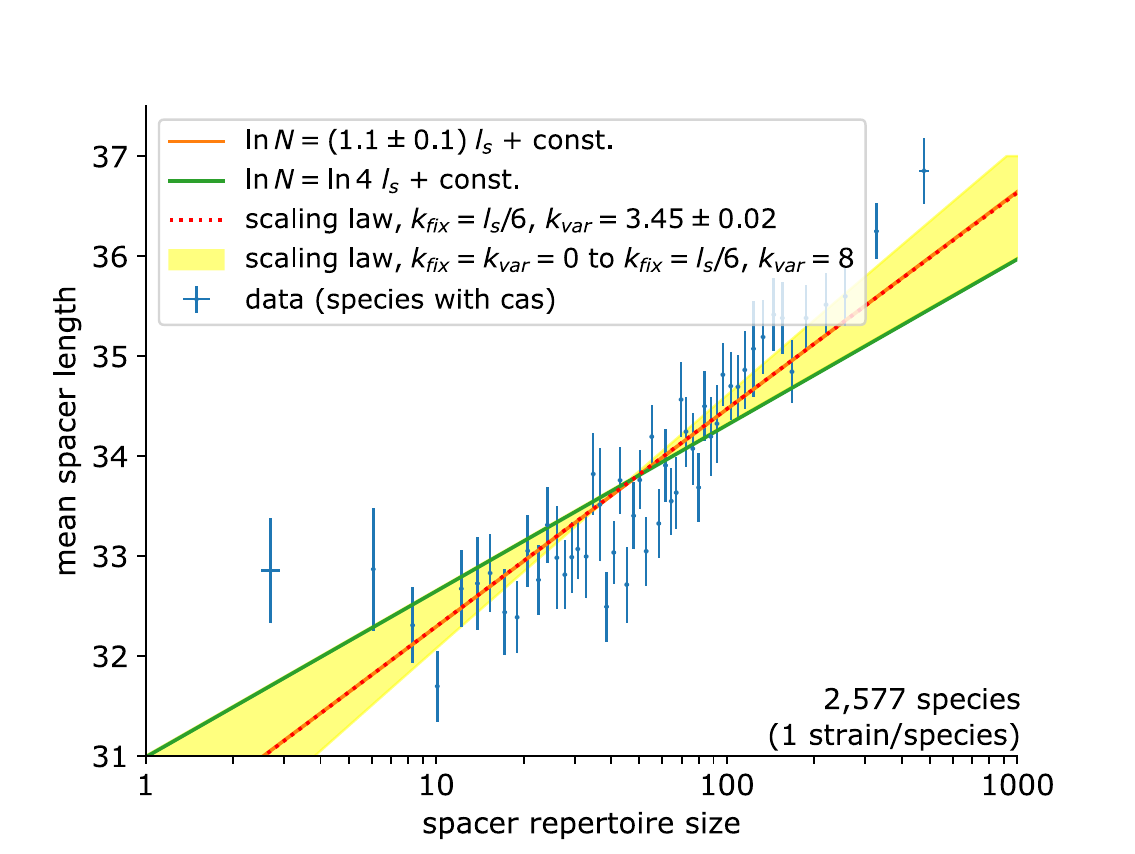}
    \caption{Cross-reactivity parameters obtained by a fit to the empirical data lie in a plausible range. The blue points are data from 2,577 species binned in increasing windows of repertoire size (50 species/bin), and the orange line is the linear fit to all species as in Fig.~\ref{fig:data}c. The green line is the naive $\ln 4$ scaling (see Eq.~\ref{eq:linear-relation}). The fitted slope is consistent with a broad range of cross-reactivity parameters (yellow region).
    A best-fit to Eq.~\ref{eq:scaling} was performed, in which $l_p$ was fixed at 3, and $k_{\text{fix}}$ was set to $l_s/6$, consistently with a 6-nt periodicity in mismatch tolerance in type I-E systems \cite{Fineran2014,Jung2017} and approximately 5 allowed mismatches in type II systems in which most spacer lengths are $\sim\! 30$ nt \cite{Fu2013,Hsu2013}.
    We found best-fit values of $k_{\text{var}} = 3.45 \pm 0.02$ and $\log_{10}(L/P) = 11.37 \pm 0.03$, where the errors are 90\% confidence intervals. The estimate of $k_{\text{var}}$ is consistent with primed acquisition tolerating many mismatches, up to 10 in some systems \cite{Fineran2014,Staals2016}, and the estimate of  $L/P$ implies a maximum risk of self-targeting $P$ in the range of $10^{-4}$ to $10^{-5}$. We expect these cross-reactivity parameters to show significant variation around these means in individual species and systems (see Fig. \ref{fig:evolution}).
    }
    \label{fig:fit_model}
\end{figure}

\begin{figure}[h]
    \includegraphics{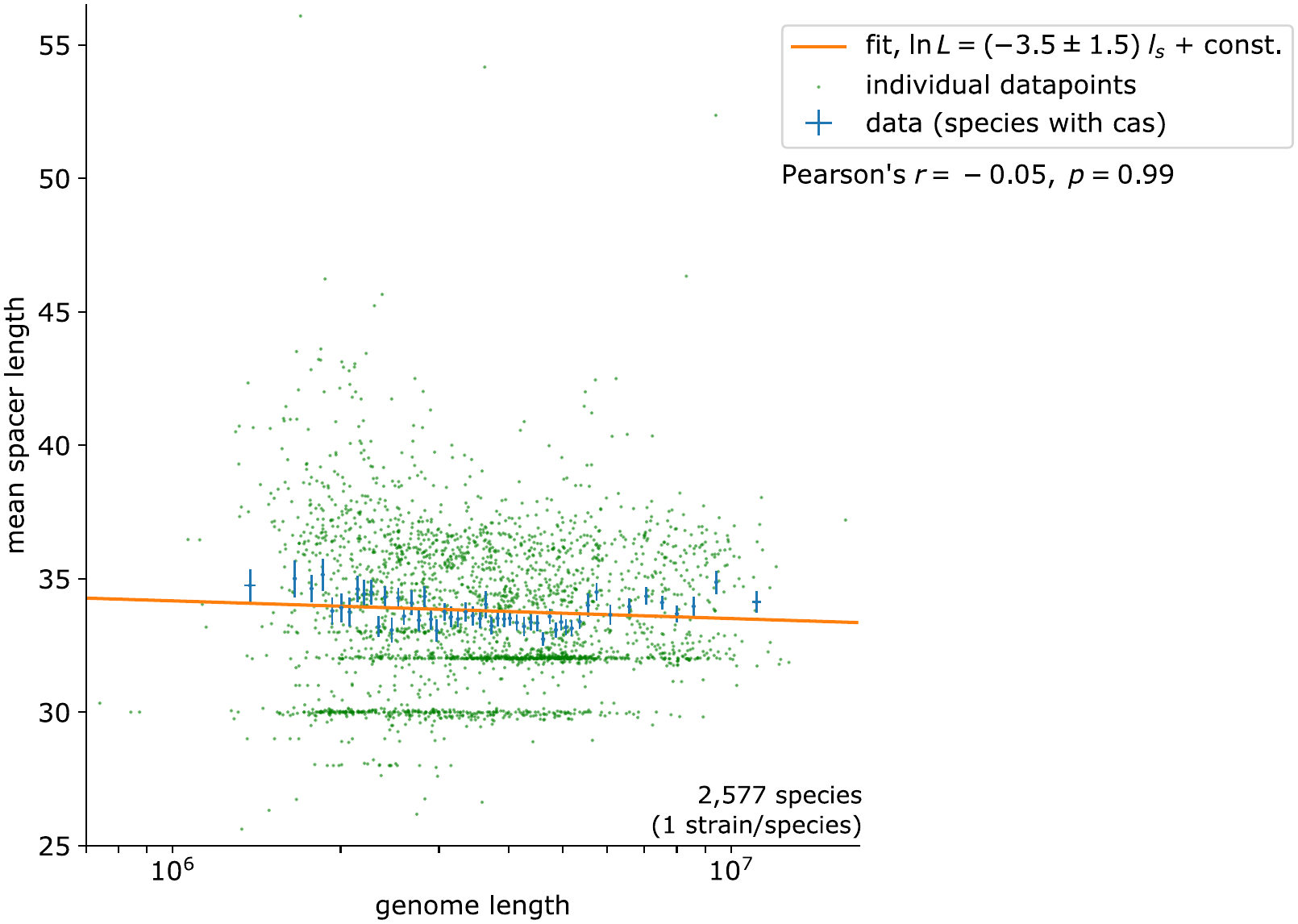}
    \caption{No correlation between genome length and mean spacer length in species with cas. The blue points are data from 2,577 species binned in increasing windows of repertoire size (50 species/bin) with error bars denoting the standard error of the mean. The green points are all 2,577 individual datapoints, and the orange line is the linear fit to the individual species. Dense horizontal bands along mean spacer length = 30 nt and 32 nt reflect many species acquiring spacers of these lengths (see Fig.~\ref{fig:data}a). Unlike Fig.~\ref{fig:data}c and Fig.~\ref{fig:fit_alldata}, no significant correlation between genome length and mean spacer length was observed.
    }
    \label{fig:fit_genomelen}
\end{figure}

\begin{figure}[h]
\includegraphics{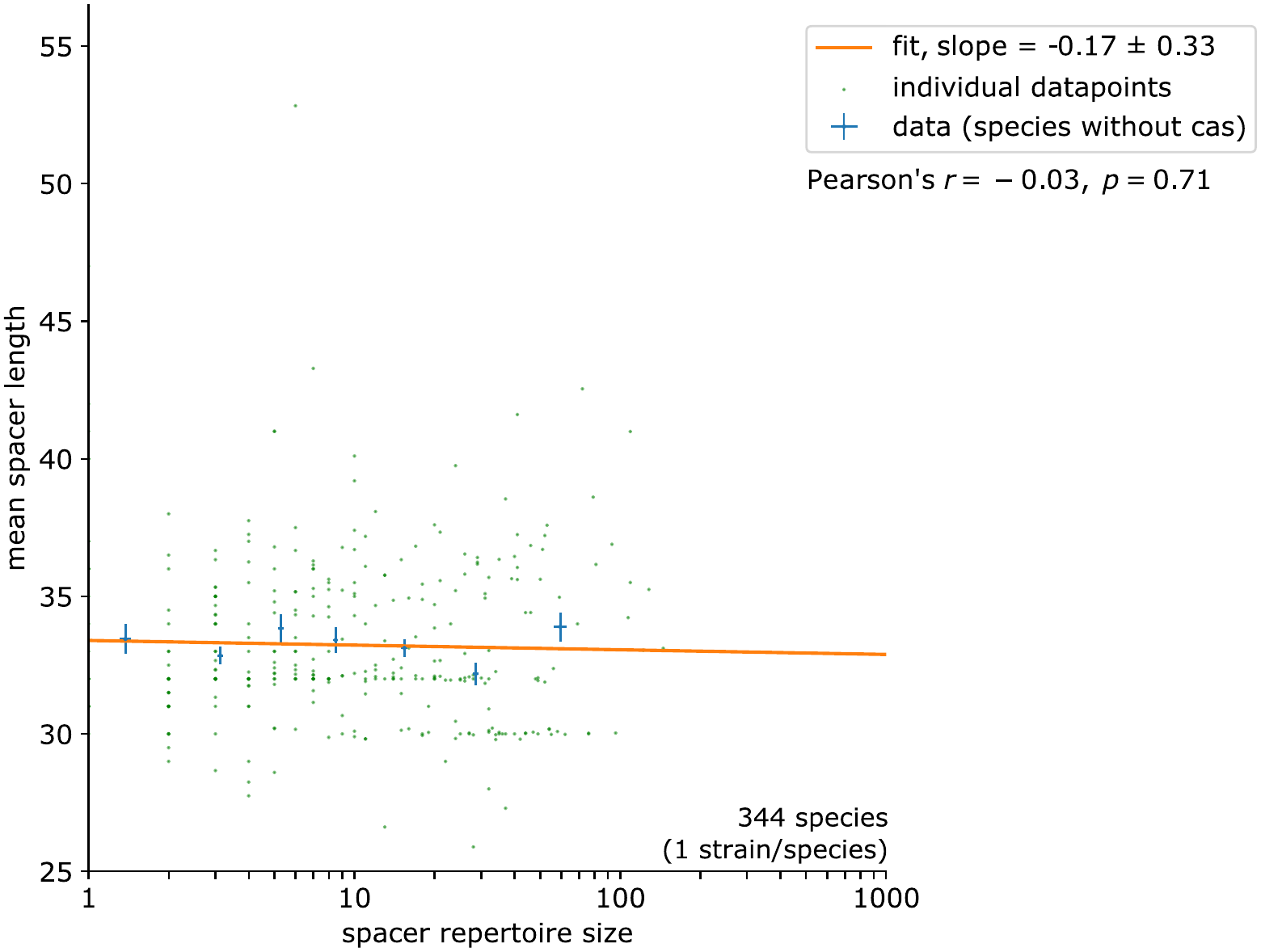}
    \caption{No scaling law between CRISPR repertoire size and mean spacer length in species without cas. The blue points are data from 344 species binned in increasing windows of repertoire size (50 species/bin) with error bars denoting the standard error of the mean. The green points are all 344 individual datapoints, and the orange line is the linear fit to the individual species as in Fig.~\ref{fig:data}e.
    }
    \label{fig:fit_alldata_nocas}
\end{figure}

\begin{figure}[ht]
    \includegraphics{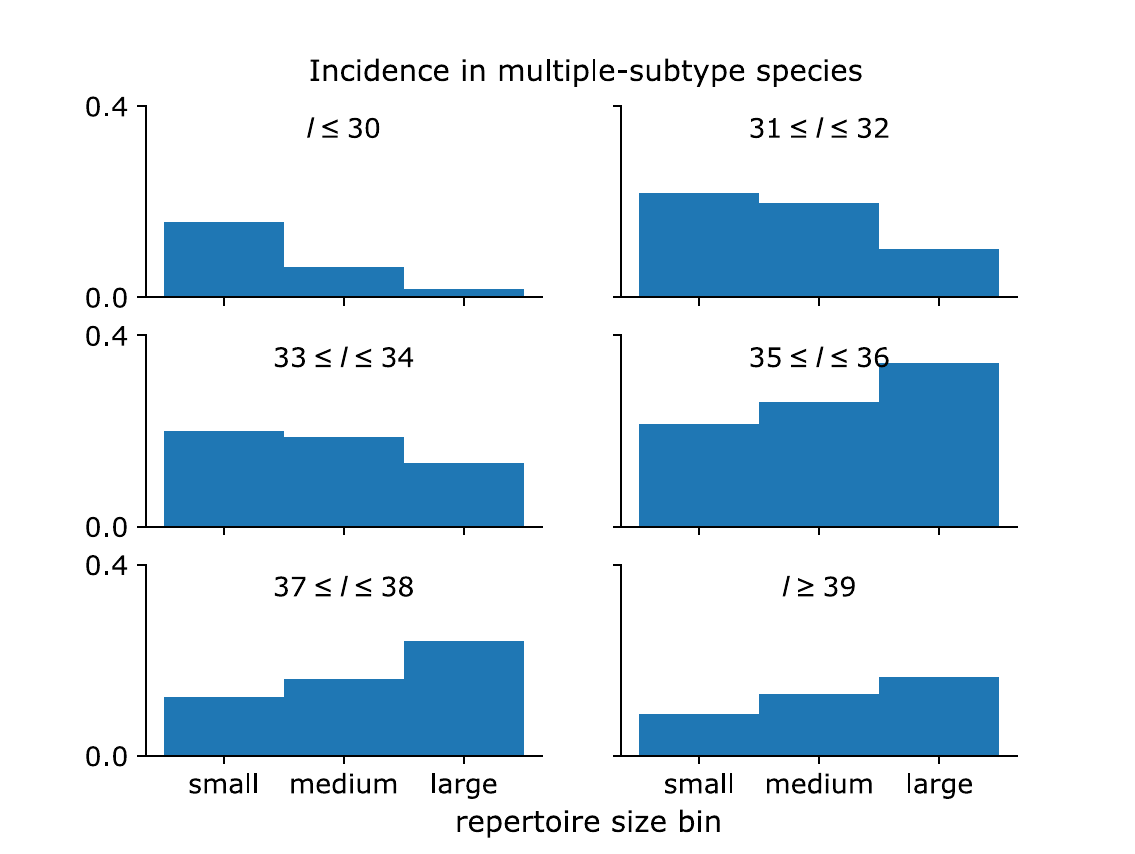}
    \caption{Variable usage of spacer lengths among species carrying multiple cas types. 843 species with multiple CRISPR-Cas systems were divided into 3 groups of 281 species having small, medium, or large repertoire sizes, respectively. All spacers were collated, and the total number of spacers in each repertoire size bin was normalized to 1, so that the bars indicate the fraction of spacers in each bin with that length. The fractional usage of spacers of length $\leq 32$ nt decreases with repertoire size, while usage of spacers of length $\geq 35$ nt increases with repertoire size among these species.}
    \label{fig:usage_l_m}
\end{figure}

\begin{figure}[h]
    \includegraphics{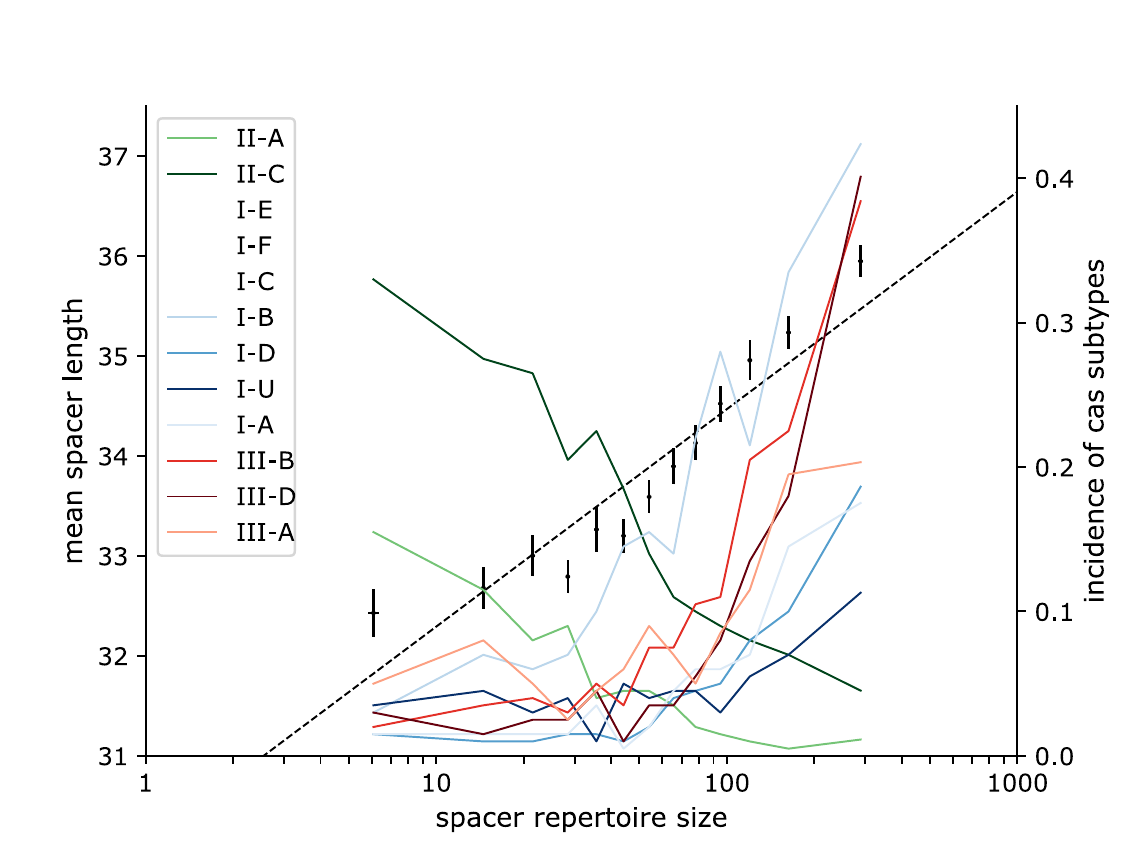}
    \caption{Variable use of cas subtypes underlies a scaling law between CRISPR repertoire size and mean spacer length in species with cas. The black points are data from 2,577 species binned in increasing windows of repertoire size (200 species/bin, 177 species in the rightmost bin) with error bars denoting the standard error of the mean. The black dotted line is the linear fit to all 2,577 individual species as in Fig.~\ref{fig:data}c. The colored lines show the subtype incidence in these same bins. The order of subtypes in the legend is given by the first quartile of spacer lengths for types II, I and III as in Fig~\ref{fig:castype}a. The incidence of types II-A and II-C clearly decreases, and the incidence of the four type I subtypes with the longest spacers and the three type III subtypes clearly increases across repertoire size bins. The incidence of the three type I subtypes with the shortest spacers do not show a clear trend and are not plotted for visibility of the other 9 subtypes. The incidence is normalized by each repertoire size bin, so the numbers indicate the fraction of species in each bin carrying that subtype (which might be a single system or one of several CRISPR-Cas systems).
    }
    \label{fig:fit-with-subtypes}
\end{figure}

\begin{figure}[ht]
    \includegraphics{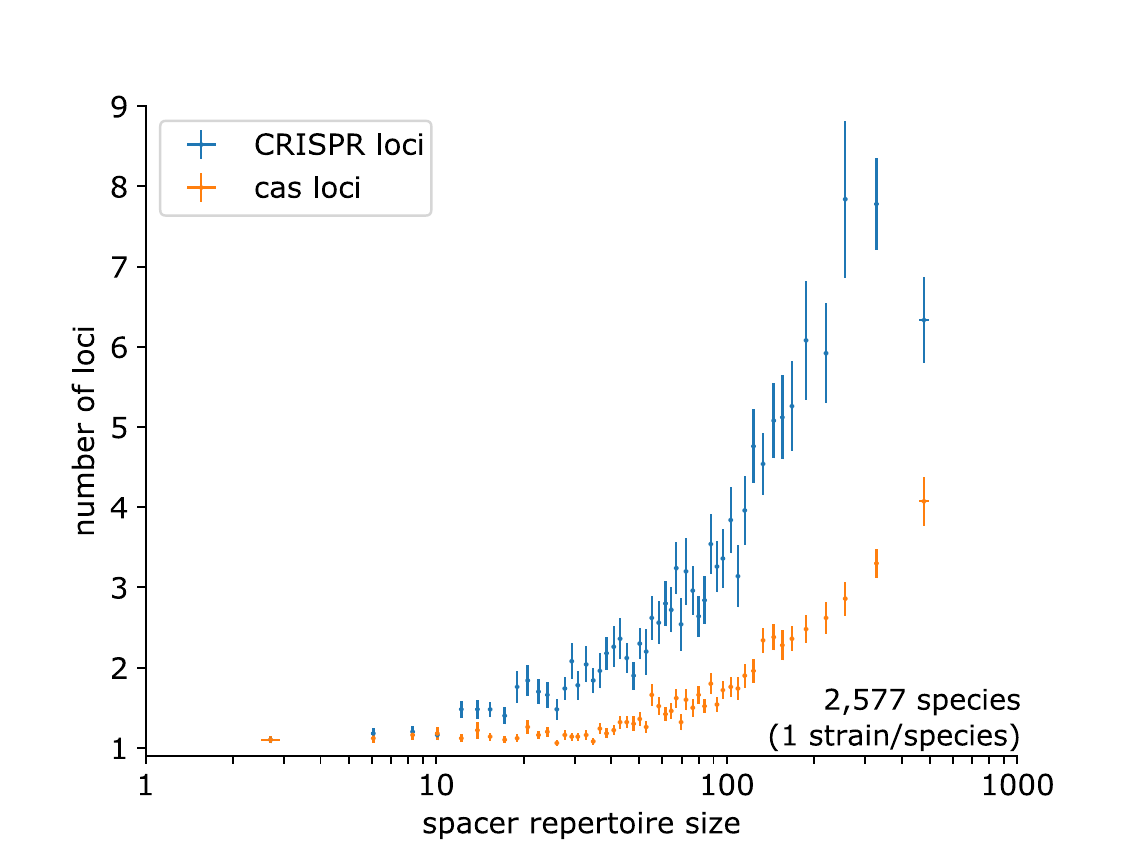}
    \caption{Spacer repertoire size is correlated with the number of CRISPR and cas loci. Data from 2,577 representative strains belonging to different species are binned by repertoire size (50 species/bin, same bins as in Fig.~\ref{fig:data}c). Error bars denote the standard error of the mean in each bin.}
    \label{fig:numloci}
\end{figure}

\begin{figure}[h]
    \includegraphics{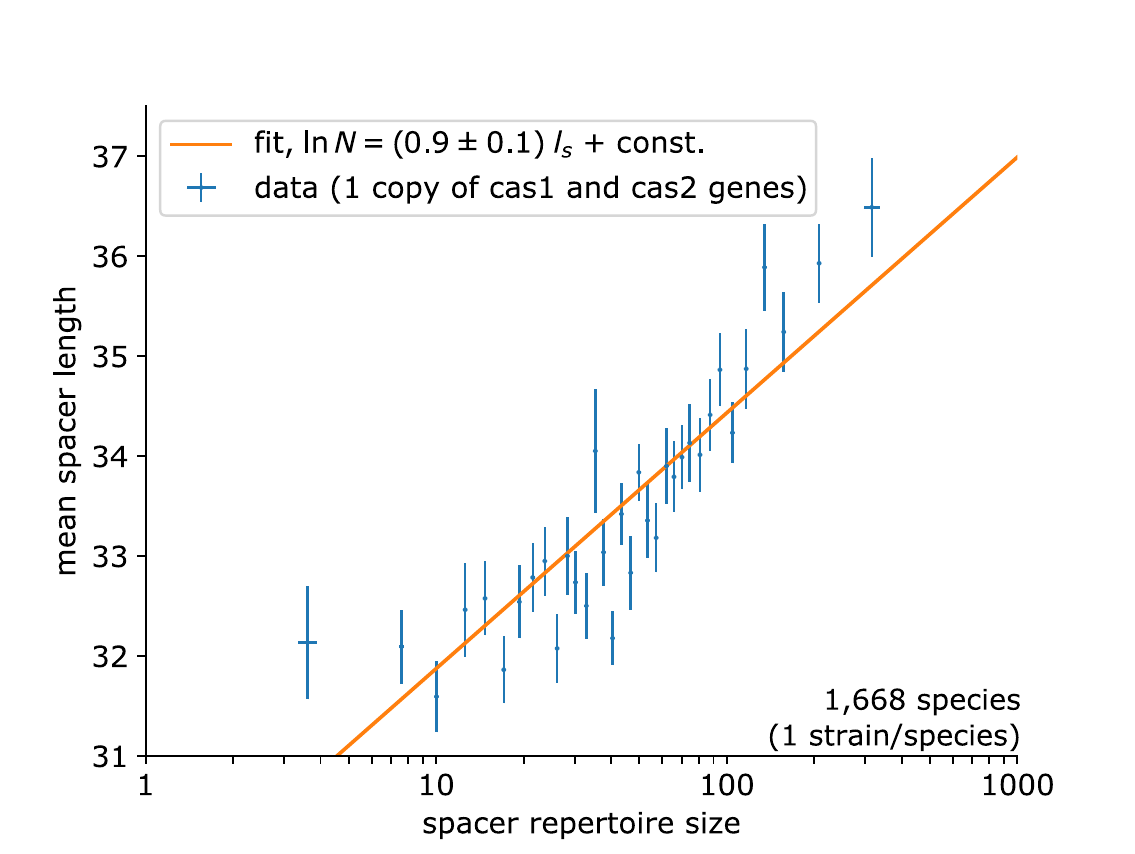}
    \caption{Repertoire size versus mean spacer length for species conditioned on one annotated gene copy of cas1 and cas2. 1,668 out of the 2,577 species contain one copy of both cas1 and cas2 genes. The orange line is a linear fit to the individual species, shown alongside the data binned by repertoire size (50 species/bin). Error bars denote the standard error of the mean in each bin.}
    \label{fig:fit_1copycas}
\end{figure}

\begin{figure}[ht]
    \includegraphics[scale=1.5]{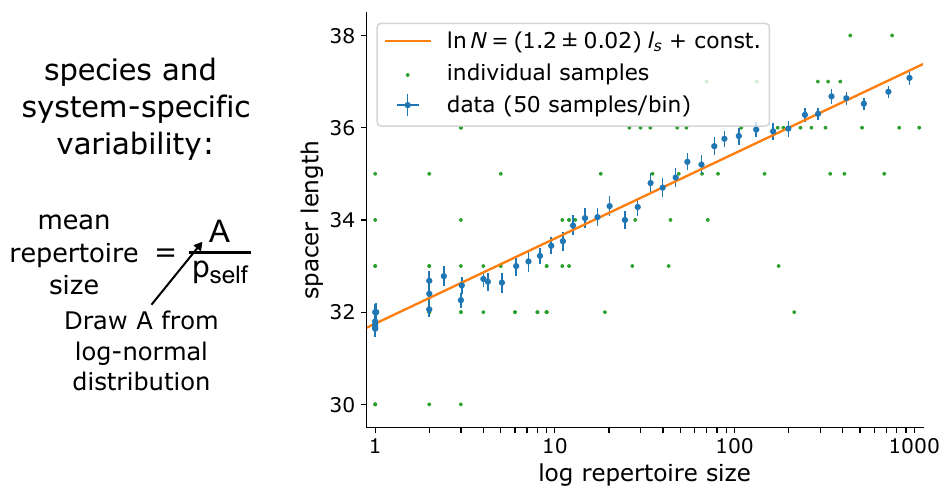}
    \caption{Species and system-specific stochasticity increases the variability of the sampled data, but a scaling law is recovered by binning by repertoire size. A sampling procedure on synthetically generated data is replicated as in Fig.~\ref{fig:evolution}. Individuals were drawn from steady-state distributions with mean proportional to $1/p_{\text{self}}$, but each time the prefactor $A$ was also drawn from a wide (log-normal) distribution with the same mean as in Fig.~\ref{fig:evolution}a--b, and standard deviation chosen such that the coefficient of variation is 1.2. A large variability in the data results, but binning recovers a clear relation between mean repertoire size and spacer length. The green points show 100 individually sampled strains, the blue points means after binning by repertoire size (50 species/bin,  2,750 species total),  and the orange line is a fit.}
    \label{fig:simulated_Alognormal}
\end{figure}

\end{document}